 \documentclass[showpacs,amsmath,amssymb,twocolumn,superscriptaddress,notitlepage,preprintnumbers,pra]{revtex4-1}

\usepackage[dvips]{graphicx}
\usepackage{amsmath,amssymb,amsthm,mathrsfs,amsfonts,dsfont}
\usepackage{amsmath,amssymb}
 
\usepackage{dcolumn}
\usepackage{bm}
\usepackage[mathlines]{lineno}
\usepackage{amsmath}
\usepackage{amssymb}
\usepackage{braket}
\usepackage{physics}
\usepackage{amsthm}
\usepackage{natbib}
\usepackage{mathtools}
\usepackage{diagbox}
\usepackage[utf8]{inputenc}
\usepackage[english]{babel}
\usepackage{scalerel}[2014/03/10]
\usepackage{stackengine}
\usepackage{algorithm}
\usepackage{relsize}
\usepackage[noend]{algpseudocode}

\usepackage{hyperref}
\hypersetup{colorlinks=true, linkcolor=blue, citecolor=blue, urlcolor=black }

\DeclareMathOperator*{\Modot}{\text{\raisebox{0.25ex}{\scalebox{0.5}{$\bigodot$}}}}

\renewcommand{\tr}{\textup{Tr}}

\algrenewcommand\algorithmicrequire{\textbf{Input:}}
\algrenewcommand\algorithmicensure{\textbf{Output:}}

\makeatletter
\newcommand{\algmargin}{\the\ALG@thistlm}
\makeatother
\newlength{\whilewidth}
\algdef{SE}[parWHILE]{parWhile}{EndparWhile}[1]
  {\parbox[t]{\dimexpr\linewidth-\algmargin}{%
     \hangindent\whilewidth\strut\algorithmicwhile\ #1\ \algorithmicdo\strut}}{\algorithmicend\ \algorithmicwhile}%
\algnewcommand{\parState}[1]{\State%
  \parbox[t]{\dimexpr\linewidth-\algmargin}{\strut #1\strut}}

\newtheorem{theorem}{Theorem}[]

\newtheorem{lemma}[theorem]{Lemma}

\theoremstyle{definition}

\newcommand{\PKU}{Center on Frontiers of Computing Studies, School of Computer Science, Peking University, Beijing 100871, China}

\newcommand{\bytedance}{ByteDance Research, Fangheng Fashion Center, No. 27, North 3rd Ring West Road, Haidian District, Beijing 100098, China}
\newcommand{\oxford}{Clarendon Laboratory, University of Oxford, Parks Road, Oxford OX1 3PU, United Kingdom}

\begin{document}

\title{Quantum computing quantum Monte Carlo algorithm}

\author{Yukun Zhang}
\affiliation{\PKU}
\author{Yifei Huang}
\thanks{Contact author: huangyifei.426@bytedance.com}
\affiliation{\bytedance}
\author{Jinzhao Sun}
\thanks{Contact author: jinzhao.sun.phys@gmail.com}
\affiliation{\oxford}

\author{Dingshun Lv}
\affiliation{\bytedance}
\author{Xiao Yuan}
\thanks{Contact author: xiaoyuan@pku.edu.cn}
\affiliation{\PKU}

\begin{abstract}
Quantum computing (QC) and quantum Monte Carlo (QMC) represent state-of-the-art quantum and classical computing methods, respectively, for understanding many-body quantum systems. 
However, straightforward integration of the two methods may encounter significant challenges, such as exponential sampling cost and inefficient walker propagation.
Here, we propose an efficient hybrid quantum-classical algorithm that integrates the two methods, overcoming these limitations while leveraging their strengths in representing and manipulating quantum states. To measure the effectiveness of the hybrid approach, we first introduce non-stoquasticity indicators (NSIs) and their theoretical upper bounds, which quantify the severity of the sign problem, a major limitation of QMC. Next, we present a hybrid QC-QMC method where the walkers are represented by quantum states prepared by a shallow quantum circuit. Although the Hamiltonian in the quantum state walker basis is not sparse, we for the first time offer an efficient and scalable approach to implement walker propagation using a quantum computer. From the QMC perspective, our algorithm significantly mitigates the sign problem in the quantum state walker basis. From the QC perspective, integrating QMC increases the expressivity of shallow quantum circuits, enabling more accurate computations that are traditionally achievable only with much deeper quantum circuits. Our method has immediate applications in tackling complex quantum many-body problems. We numerically test and verify it for the N$_2$ molecule (12 qubits) and the Hubbard model (16 qubits), observing a significant suppression of the sign problem (which exponentially decreases with circuit depth) and a notable improvement in calculation accuracy (which is about two to three orders compared to variational quantum algorithms). Our work paves the way to solving practical problems with intermediate-scale and early-fault tolerant quantum computers, with broad applications in chemistry, condensed matter physics, and materials.


 
 
\end{abstract}

\maketitle

\section{Introduction}\label{sec:intro}

A critical task in quantum physics is finding eigenstates and eigenenergies of quantum many-body systems due to their wide-ranging applications in quantum chemistry~\cite{chan2011density,mcardle2020quantum,cao2018quantum}, materials~\cite{Bauer2020}, condensed matter physics~\cite{zheng2017stripe}, high energy physics~\cite{Jordan_2012,PhysRevLett.126.062001}, etc.
Various classical computational methods have been developed based on different approximations, including perturbation theories~\cite{abrikosov2012methods}, variational tensor network approaches~\cite{WFT1,orus2019tensor,jordan2008classical}, self-consistent density functional theory or embedding methods~\cite{DFT2,knizia2012density,rubin2016hybrid,li2021toward,Troyer16}, quantum Monte Carlo (QMC)~\cite{ceperley1980ground,zhang1997constrained,booth2009fermion}, machine learning~\cite{carleo2017solving,hermann2020deep,pfau2020ab}, etc.
{Among these} state-of-the-art classical methods, QMC has drawn great attention and has been widely exploited to study problems in chemistry and condensed matter physics, solving systems with up to around 800 electrons~\cite{al2021interactions}. 
{However, QMC has its limitations in addressing the eigenstate problem}.
Specifically, it represents quantum states by an effective superposition of classical basis states, which generally suffers from the notorious sign problem~\cite{troyer2005computational}, prohibiting its effectiveness for strongly correlated systems.


{
Quantum computers can naturally represent quantum states and have the potential to overcome the aforementioned challenges~\cite{nielsen2002quantum}. Quantum algorithms, such as adiabatic state preparation~\cite{albash2018adiabatic}, quantum phase estimation (QPE)~\cite{kitaev1995quantum,abrams1999quantum,aspuru2005simulated}, and the more recent ones based on spectral filters~\cite{lin2022heisenberg,wan2022randomized,zeng2021universal,huo2021shallow,dong2022ground,Ge19,lu2021algorithms,sun2024high}, have been proposed to solve the eigenstate problem. Nevertheless, those quantum algorithms generally require a deep quantum circuit, which is only realizable with fault-tolerant universal quantum computers. 
For example, even the simplest Ising toy model of fewer than 100 qubits needs {billions} of quantum gates, presenting a formidable resource requirement for current quantum technology~\cite{sun2024high}.
Furthermore, these algorithms have their assumptions, such as a nonvanishing energy gap for nearly all the existing quantum algorithms and a nonvanishing initial state overlap for spectral filter methods and QPE, whose validity is not justified in general. For near-term applications, a long-standing question is whether shallow circuits can be useful for solving the eigenstate problem. 
Although various variational quantum algorithms have been developed for this task~\cite{peruzzo2014variational,cerezo2021variational,RevModPhys.94.015004}, the answer is not straightforward due to several challenges.
On the one hand, near-term quantum computers can only realize shallow quantum circuits, which generally have limited expressivity and capacity owing to the small circuit depth~\cite{PhysRevLett.128.080506}. On the other hand, if the ansatz is too deep, the optimization becomes harder~\cite{anschuetz2022beyond}, the gradients of the parameters might vanish exponentially~\cite{mcclean2018barren}, and the quantum state would accumulate too much noise to be classically simulable~\cite{schuster2024polynomialtimeclassicalalgorithmnoisy}. 
These challenges are inter-correlated, making it an open question whether near-term quantum computers with shallow circuits can be effectively utilized to solve eigenstate problems.}

As state-of-the-art quantum and classical computing methods, QC and QMC each have their respective strengths and limitations. A recent emerging field explores the intersection of these two research directions. 
A pioneering work by the Google team~\cite{Huggins_2022} proposed to replace the trial wave function in the auxiliary field (AF)-QMC with a quantum state to ``unbias'' the computation. This also aligns with the ideas previously studied in Refs.~\cite{wouters2014projector,clark2014stochastically}, which consider complicated classical trial states. 
However, Mazzola and Carleo~\cite{mazzola2022exponential} pointed out a potential challenge in Google's work, namely, the difficulty in estimating overlaps between the trial wave function and the classical walkers, which generally require an exponential number of samples as the system scales up.
Moreover, other potential issues with Google's work~\cite{Huggins_2022} are the sign problem and the bias. Specifically, since the walkers are classical, the sign problem and the bias remain unresolved unless the trial wave function has the same sign structure as the target ground state. 
Consequently, as noted by the Google team~\cite{lee2022response}, their method remains a heuristic strategy, which applies efficiently to problems with intermediate scales. 
{Recently, Xu and Li~\cite{xu2023quantum} introduced an alternative approach to exploiting the denominator computed by the classical wavefunction to further mitigate the sampling overhead. However, this is still a heuristic method, which cannot cure the exponential challenge generally unless the trial wavefunction and the classical wavefunction have the same amplitude for the exponential configurations.}


In existing works, a quantum computer is used to prepare the trial wave function, while the walkers are represented and propagated classically. A potentially more effective approach is to use quantum computers to represent the walkers directly, which may lead to faster convergence and reduced sign problems. The concept of using walkers in a rotated \emph{classical} basis has been explored in different contexts~\cite{sandvik2010loop,PhysRevLett.126.216401}, showing clear numerical improvements. However, since the rotated basis remains classical and cannot accurately approximate the eigenstates of strongly correlated systems, these methods still suffer from severe sign problems in general. In contrast, using quantum computers to prepare the walkers provides a more powerful means of approximating many-body quantum states, offering greater potential for mitigating the sign problem. However, a significant challenge arises in propagating general quantum state walkers. Specifically, the Hamiltonian becomes non-sparse in the quantum state walker basis, necessitating exponential walker propagation costs (gate and measurement complexity). Thus, enhancing QMC with quantum state walkers remains elusive.


\begin{figure*}
  \includegraphics[width=0.98\textwidth]{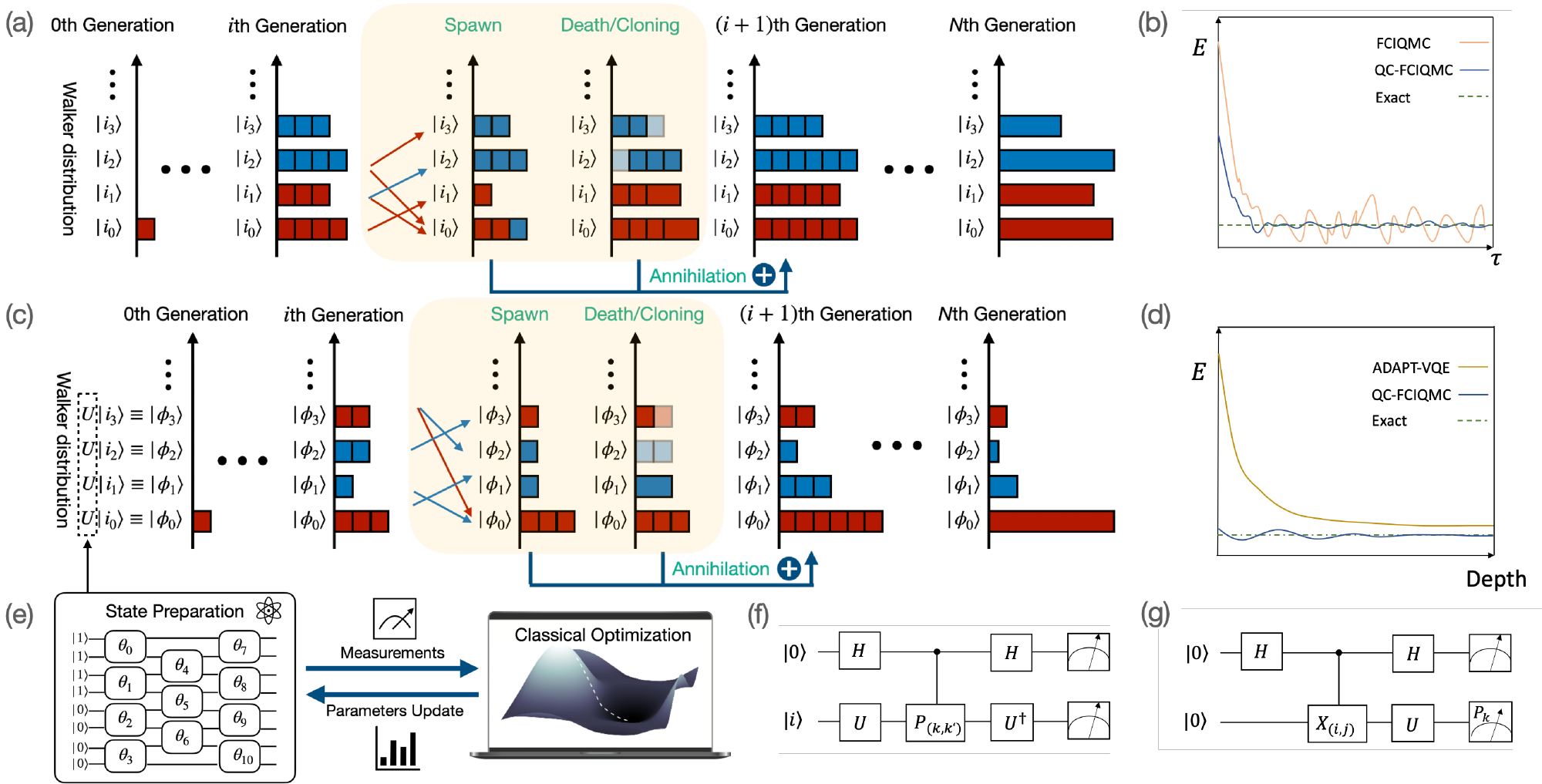}
  \caption{(a) Sketch of the spawning, death/cloning, and annihilation steps in FCIQMC. Red and blue colors represent walkers with positive and negative signs respectively. Red and blue arrows represent the sign of $H_{ji}$. The transparent boxes represent dead walkers, while the elongated walkers are the cloned ones. (b) {A sketched cartoon of the expected sign problem comparison for our QC-FCIQMC and FCIQMC.} (c) The procedure of QC-FCIQMC. The quantum circuit $U$ that generates the new basis is obtained from VQAs, such as the one shown in (e). (d) {A sketched cartoon of the expected energy convergence comparison of ADAPT-VQE and QC-FCIQMC.} (f) Key quantum circuits for realizing walker propagation. Here $\ket{\phi_i} = U\ket{i}$ is the quantum state walker with the unitary $U$, the controlled-$P_{(k,k')}$ gate denotes $\ket{0}\bra{0}\otimes P_k + \ket{1}\bra{1}\otimes P_{k'}$ with $P_k$ being Pauli matrices. When the Hamiltonian has the decomposition $H=\sum_k h_kP_k$, the quantum circuit evaluates/samples according to $p_{k,k'}^i(j)=\mathrm{Re}\langle{i|U^\dag P_k U\Pi_j U^\dag P_{k'}U|i}\rangle$, which constitute $\abs{\widetilde{H_{ji}}}^2=\sum_{k,k}h_kh_{k'}p_{k,k'}^i(j)$, here $\Pi_j=\ket{j}\bra{j}$.We refer to the main text for the protocol detailing how to use the quantum circuit to achieve the propagation of quantum state walkers. (g) Circuit for evaluating the real part of $\widetilde{H_{ji}}$. The imaginary part could be similarly obtained by adding an $S$ gate before the last $H$ gate. The circuit is used to track the sign/phase during walker propagation.}\label{FRAMEWORK}
\end{figure*}

Here, we introduce a hybrid approach that integrates QC and quantum QMC, leveraging their complementary strengths in representing and processing quantum states while mitigating their respective weaknesses—the limitations of shallow circuits for QC and the sign problem for QMC. We utilize walkers prepared by a shallow quantum circuit and, for the first time, develop a novel method for efficiently implementing walker propagation using quantum computers. This approach addresses the aforementioned challenges and offers a theoretically innovative and practically efficient way to combine QC and QMC.
The structure of the paper is organized as follows. 
After reviewing the background of QMC and QC in Sec.~\ref{sec:bg}, we introduce non-stoquasticity indicators (NSIs) and their upper bounds in Sec.~\ref{sec:sign_problem}, which measure the seriousness of the sign problem. We then consider full configuration interaction quantum Monte Carlo (FCIQMC)~\cite{booth2009fermion, cleland2010communications} and introduce QC-FCIQMC with quantum states (walkers) prepared with an optimized quantum circuit in Sec.~\ref{Sec:QCFCIQMC}. We propose an efficient way to scalably implement the QC-FCIQMC procedure by generating walker propagation trajectories using quantum circuits. 
{We demonstrate that the sampling procedures and the post-processing procedures are efficient and independent of the system size due to a deliberate use of quantum measurements.}
Since the walker basis is optimized, the sign problem of FCIQMC is greatly alleviated, as verified numerically for the N$_2$ molecule (12 qubits) and the Hubbard model (16 qubits) in Sec.~\ref{Sec:numerics}. We show a drastic decrease in NSIs, the energy variance (with a fixed number of walkers), and the number of walkers required to ensure a desired variance using our method. The results also show much higher accuracies for QC-FCIQMC, achievable only with much deeper quantum circuits in conventional quantum computing approaches.  

\section{Background}\label{sec:bg}



We begin by covering the basics of quantum Monte Carlo (QMC) and quantum computing (QC) in finding the ground state of Hamiltonian $H$, setting the stage for our work. We discuss the limitations of these two approaches and explain how their combination can potentially overcome these limitations. 
In the main text, we focus on $H$ with real matrix elements, while the general case is discussed in Supplementary Information~\cite{notesupp}.

We consider projector QMC, a subclass of QMC, which effectively represents the state as a superposition of walkers (states such as Slater determinants) and realizes the imaginary-time evolution (ITE) of the state by stochastically propagating the walkers. Consider Full configuration interaction quantum Monte Carlo~(FCIQMC) as an explicit example, denote the walkers by $\{\ket{i}\}$ and the initial state by 
$
\ket{\psi(0)}=\sum_{i}c_i(0)\ket{i}   
$ with coefficients $c_i(0)$, FCIQMC aims to realize the ITE of the state as
\begin{equation}
    \ket{\psi(\tau)}\propto e^{-(H-S)\tau}\ket{\psi(0)}
\end{equation}
with imaginary time $\tau>0$ and a certain parameter $S$. 
Specifically, for FCIQMC (see Fig.~\ref{FRAMEWORK}(a)), we first generate $N_i(\tau)$ number of walker $\ket{i}$ at $\tau=0$; Then, for small $\Delta \tau$, we repeatedly update each walker $\ket{i}$ through iterating the following procedures. 
\begin{enumerate}
    \item Spawning~---~spawn a child walker $\ket{j}$~($j\neq i$) with probability $\abs{H_{ji}}\Delta \tau$ with the same sign as walker $\ket{i}$ multiplied by  $-H_{ji}/|H_{ji}|$ ($H_{ij}=\langle i|H|j\rangle$); 
    \item Death or cloning~---~the walker $\ket{i}$ dies with probability $(H_{ii}-S)\Delta \tau$ (if $H_{ii}-S>0$) and clones itself with probability $\abs{(H_{ii}-S)}\Delta \tau$ otherwise; 
    \item Annihilation~---~annihilate same walker pairs with opposite signs.  
\end{enumerate}
For problems with complex coefficients, we separately generate and evolve walkers for the real and imaginary parts. 
While QMC could deterministically find the ground state with large $\tau$, it suffers from the sign problem~\cite{troyer2005computational}. 
Specifically, since the walkers may have both positive and negative signs, the average sign generally decreases exponentially with time $\tau$ and the system size. 
We will shortly present a quantitative measure of the sign problem. In general, the sign problem is difficult to quantify precisely. However, we derive an upper bound on this measure, which is computationally efficient to calculate. This result may be of independent interest, as the upper bound could serve as a general cost function for mitigating the sign problem in QMC methods under walker basis ratations~\cite{hangleiter2020easing,levy2021mitigating}. We further present our hybrid QC-FCIQMC method, which adopts quantum state walkers to mitigate the sign problem.\\

For quantum computing, we mainly focus on variational quantum algorithms (VQA) tailored for near-term quantum devices (see Fig.~\ref{FRAMEWORK}(e) for example). 
Here, we use the variational quantum eigensolver (VQE)~\cite{peruzzo2014variational,grimsley2019adaptive} as a representative example. 
For the Hamiltonian $H$, we approximate its ground state as  $|{\phi_0(\vec \theta)}\rangle = U(\vec \theta)\ket{\bar 0}$ with a parameterized quantum circuit $U(\vec \theta)$ and an easy-to-prepare initial state $\ket{\bar 0}$. Then, we employ a classical optimizer to search for parameters $\vec \theta$ that minimize the energy $E(\vec \theta) = \langle\phi_0(\vec \theta)|H|\phi_0(\vec \theta)\rangle$. Based on the variational principle, the solution to the minimization problem would provide an upper bound to the ground state energy. Similar concepts have been widely applied in classical ab initio calculations in chemistry, materials, and condensed matter physics~\cite{WFT1,orus2019tensor,jordan2008classical}. However, since quantum computers could represent quantum states and manipulate quantum entanglement more efficiently, they potentially enable VQE to provide more accurate solutions than classical methods for certain problems with strong correlations.

Recent studies have demonstrated the potential power of variational quantum algorithms both theoretically and experimentally~\cite{preskill2018quantum,PRXQuantum.2.017003,endoreview,cerezo2020variational,RevModPhys.94.015004,kandala2017hardware,google2020hartree,cao2023generation,guo2024experimental}. However, VQAs face several theoretical and practical challenges.  First, the limited coherence time of near-term quantum devices restricts the circuit depth, which may impede the approximation of the desired ground state~\cite{Abbas2021,holmes2021connecting}. 
Additionally, noise in quantum gates degrades the computation accuracy; deeper quantum circuits introduce more noise, making it difficult to surpass classical methods~\cite{PhysRevX.7.021050,PhysRevLett.119.180509,PhysRevX.8.031027,mitigationIBM22}. 
Furthermore, the energy landscape $E(\vec \theta)$ with deeper quantum circuits may have more local minima~\cite{anschuetz2022beyond}, complicating the convergence to the optimal solution.
Meanwhile, deep (too expressive) circuits can also suffer from
barren plateau problems~\cite{cerezo2020impact,Wang_2021,PhysRevLett.127.120502}, where gradients vanish exponentially with the system size. 
Lastly, the lack of automatic differentiation complicates efficient optimization, especially for deep quantum circuits with numerous parameters. 
We can see that these limitations are intrinsically linked to circuit depth. On one hand, deeper quantum circuits are necessary to enhance expressivity; on the other hand, they introduce issues like noise, local minima, barren plateaus, and optimization difficulties. Thus, balancing these limitations is crucial for the success of VQAs. 

Below, we show that the combination of QMC and VQA  could circumvent the above challenges. By running a shallow quantum circuit (which experiences less severe issues due to noise, local minima, barren plateaus, and optimization) and leveraging QMC to improve the circuit's expressivity, we can create a more effective method that optimally utilizes current quantum hardware. Theoretically, it has been demonstrated that shallow quantum circuits can generate quantum correlations that classical computers cannot efficiently emulate~\cite{arute2019quantum}. While these complex correlations alone may not suffice to represent the ground state of a specific problem, they can exhibit high expressivity when combined with QMC. In other words, the shallow quantum circuit provides a resource of states with intricate correlations, and QMC further refines these states to better approximate the ground state of a target Hamiltonian. For a more detailed description of our method, see Sec.~\ref{Sec:QCFCIQMC}, and for a numerical verification of our argument, see Sec.~\ref{Sec:numerics}. 




%

\section{Sign problem indicators}\label{sec:sign_problem}
Before introducing our method, we present a new measure for quantifying the sign problem and prove computable upper bounds for the measure. These results provide the theoretical basis for why using a quantum state walker in QMC could alleviate the sign problem, as also numerically verified in Sec.~\ref{Sec:numerics}. It also serves as an independent tool for quantifying the sign problem and has potential applications in the classical optimization of the rotation basis in QMC.

As pointed out by Troyer~\cite{troyer2005computational}, a necessary condition for the sign problem is non-stoquasticity. Consider a general walker basis (orthonormal states) $\{\ket{\phi_i}\}$,
a Hamiltonian is called ``stoquastic'' when all its off-diagonal terms (in the walker basis) are non-positive.
For a thermal state $e^{-\beta H}/Z$ with $Z=\tr[e^{-\beta H}]$, the expectation of observable $A$ is
$\langle A\rangle = \tr[A e^{-\beta H}]/Z$. Denote $G = \alpha I-H$ with $\alpha=\max_{i}H_{ii}$, and we have
\begin{equation}\label{Eq:Aexpansion}
    \langle A\rangle = \frac{\tr[Ae^{\beta G}]}{\tr[e^{\beta G}]} = \frac{\sum_{k=0}^{\infty}\frac{\beta^k}{k!}\sum_{i_0,i_1,\dots i_{k}}A_{i_0i_1}G_{i_1i_2}\cdots G_{i_{k}i_0}}{\sum_{k=0}^{\infty}\frac{\beta^k}{k!}\sum_{i_1,i_2\dots i_{k}}G_{i_1i_2}G_{i_2i_3} \cdots G_{i_{k}i_1}},
\end{equation}
where 
we denote 
$M_{ij} := \langle{\phi_i|M|\phi_j}\rangle$ for any operator $M$.
When $H$ is stoquastic, every element of $G$ is positive, and so are the expanded terms of $\tr[e^{\beta G}]$. 
On the other hand, when $H_{ij}$ has positive off-diagonal terms, the expansion has a mixture of signs, leading to the sign problem. 

To quantify the sign problem of Hamiltonian $H$, {we define an indicator for the sign problem, which is termed non-stoquasticity indicator (NSI).}
We partition the Hamiltonian as
$H = H_++H_-$ with $(H_-)_{ij} = H_{ij}~([i=j]~or~[i\neq j~and~H_{ij}<0])$ 
and $(H_+)_{ij} = H_{ij}~[i\neq j~and~H_{ij}>0]$. The bosonic
form~\cite{troyer2005computational} of $H$ is $\tilde H=H_--H_+$, which is stoquastic and hence has no sign problem. The NSI is defined by the normalized difference between $H$ and its bosonic form $\tilde H$ as
\begin{equation}
    S(H) = \frac{\textup{Tr}[e^{-\beta \tilde H}]-\textup{Tr}[e^{-\beta H}] }{\textup{Tr}[e^{-\beta H}]}.
\end{equation}
The NSI is closely related to the average sign of the expansion. 
Specifically, denote $s$ to be the sign of each term in the expansion of $\tr[e^{\beta G}]$ in Eq.~\eqref{Eq:Aexpansion}, then we have $S(H)=1/\langle{s}\rangle-1$. The sign problem indicates an exponentially small average sign $\langle{s}\rangle = e^{-\beta \Delta f}$ with $\Delta f$ being the free energy difference between $H$ and $\tilde H$~\cite{troyer2005computational}.  Thus $S(H)$ also exponentially increases with $\beta\Delta f$. However, it is generally hard to evaluate $\Delta f$, $\langle{s}\rangle$, or $S(H)$. To address this problem, we derive a theoretical upper bound of $S(H)$ independent of $\Delta f$.
 

\begin{theorem}
\label{theorem:NSI_upper_bound}
The NSI, $S(H)$, is upper-bounded by
\begin{equation}
S(H)\le 2e^{\beta \|(\alpha - H_-)\|_{L_1}}\sinh(\beta\|H_+\|_{L_1}),
\end{equation}
where  $\|M\|_{L_1} := \sum_{i,j}|M_{ij}|$ for matrix $M$.
\end{theorem}
\noindent  For a stoquastic Hamiltonian, we have $H_+=0$, resulting in $S(H)=0$, which indicates the absence of a sign problem. This makes the upper bound tight for stoquastic Hamiltonians. Generally, the upper bound also suggests that a smaller $\|H_+\|_{L_1}$ corresponds to a less severe sign problem, which is an independent result aligning with recent studies~\cite{signproblem20,hangleiter2020easing,PhysRevResearch.2.032060,PhysRevLett.126.216401}. \\

{
In \autoref{theorem:NSI_upper_bound}, the upper bound applies to the sign problem for thermal state preparation in a general walker basis. A more interesting question is whether we can derive the NSI and its upper bound when considering the ground state with a specific input initial state. To that end, we derive a state-dependent NSI.}
Considering the imaginary time evolution of a specific initial state, $\phi_0$, with time $\tau=\beta/2$, the state-dependent NSI can be similarly defined as
\begin{equation}
    S(H,\phi_0) = \frac{\langle\phi_0|e^{-\beta \tilde H}|\phi_0\rangle - \langle\phi_0|e^{-\beta H}|\phi_0\rangle}{\langle\phi_0|e^{-\beta  H}|\phi_0\rangle}.
\end{equation}
Again, $S(H,\phi_0)$ measures the sign problem and is related to the average sign when applying imaginary time to the initial state $\phi_0$, i.e., $e^{-\tau H}\ket{\phi_0}$. An upper bound of $S(H,\phi_0)$ is given here. 
\begin{theorem}\label{thm:nsi_init}
The state-dependent NSI, $S(H,\phi_0)$, is upper-bounded by
\begin{equation}
S(H,\phi_0) =\mathcal O\left(\|\Pi_{\perp}H\ket{\phi_0}\|\right),
\end{equation}
where $\|\ket{v}\|=\sqrt{\bra{v}v\rangle}$ and $\Pi_{\perp}=I-\ket{\phi_0}\bra{\phi_0}$.
\end{theorem}
\noindent  
Here, we have ignored the dependence on other matrix elements of $H$ and refer to Supplementary Information~\cite{notesupp} for the complete results~\cite{notebounds}. 
Compared to Theorem~\ref{theorem:NSI_upper_bound}, the upper bound for state-dependent NSI is only linearly dependent on $\|\Pi_{\perp}H\ket{\psi_0}\|$. On one hand, this linear convergence speed is slower than the $\sinh$ function. On the other hand, the state-dependent NSI upper bound only relies on the quality of the input initial state, which is a much weaker requirement than minimizing $\|H_+\|_L$ in Theorem~\ref{theorem:NSI_upper_bound} (which generally requires approximately diagonalization of the Hamiltonian). Therefore, instead of minimizing $\|H_+\|_L$ or approximately diagonalizing $H$, we can aim to find a good initial state to mitigate the sign problem, which is a much easier task in practice. 

Below, we show how to use quantum computing to generate the initial state to mitigate the sign problem efficiently. In Sec.~\ref{Sec:numerics}, we also numerically simulate our algorithm by choosing a set of optimized initial states with different circuit depths. We observe that the state-dependent NSI can indeed be significantly reduced, and it exponentially decreases with the circuit depth.


%

\section{QC-FCIQMC}\label{Sec:QCFCIQMC}
Now, we introduce our hybrid quantum computing-full configuration interaction quantum Monte Carlo (QC-FCIQMC) method. As we discuss in the Introduction, unlike prior works such as Goolge's one~\cite{Huggins_2022} that utilize quantum trial wave functions, we replace the entire classical walker states $\{\ket{i}\}$ with quantum states $\{\ket{\phi_i}\}$ prepared by a quantum circuit $\ket{\phi_i} = U\ket{i}$, as shown in Fig.~\ref{FRAMEWORK}(c). These new walkers $\{\ket{\phi_i}\}$ are orthogonal by construction since $\bra{j}U^{\dagger}U\ket{i}=\delta_{ij}$. In the Heisenberg picture, this is equivalent to considering classical walkers $\{\ket{i}\}$ with a similarity transformed Hamiltonian $U^\dag H U$. As noted in the Introduction, similar concepts have been explored in the QMC community~\cite{sandvik2010loop,PhysRevLett.126.216401}, where $U$ is an efficiently computable unitary, such as matrix product operators. Extending the idea to consider general quantum circuit unitaries $U$ is a natural progression, which can further enhance the expressivity of the basis rotation and mitigate the sign problem. However, as we highlight below, efficiently realizing walker propagation (in the quantum unitary $U$ basis) poses a critical challenge, which our work addresses for the first time.

For the choice of the quantum circuit $U$, we may use a quantum computer to find $U$ that approximately diagonalizes $H$~\cite{PhysRevResearch.1.033062,PhysRevLett.122.230401}, aiming to suppress the off-diagonal elements $U^\dag H U$~\cite{UHUnote1}. One might question the need for QMC if we can already diagonalize $H$. The rationale is that the exact diagonalization of $H$ might need a deep quantum circuit, which is beyond the capability of near-term quantum devices. Contrarily, we only need to approximately diagonalize $H$ with a shallower circuit, while leveraging QMC to further improve the accuracy. 
However, approximate diagonalization remains a challenging task for current quantum computing hardware. 
Fortunately, according to Theorem~\ref{thm:nsi_init}, finding a good initial state with a significant overlap with the ground state can already help mitigate the sign problem.
Thus, instead of striving for approximate diagonalization, we can aim to find an approximate ground state, a simpler task more suitable for shallow quantum circuits~\cite{grimsley2019adaptive,zhang2020mutual,tang2019qubit,tang2021qubit2,fan2021circuit}. 
We can implement VQAs such as VQE with a shallow quantum circuit, which transforms the basis composed of single determinants into one that better aligns with the Hamiltonian eigenstates (especially the ground state). 
This quantum circuit then serves as the walker basis for QMC, which helps alleviate the sign problem (see Fig.~\ref{FRAMEWORK}(b)). 
From the quantum computing perspective, this approach allows for more accurate results with shallow quantum circuits. This reduces the need for deep quantum circuits required by conventional quantum algorithms and enhances the practical capabilities of near-term quantum computing devices (see Fig.~\ref{FRAMEWORK}(d)).\\


Once the quantum circuit $U$ is determined, for instance through VQE, we can replace the `classical' walkers $\ket{i}$ with `quantum' walkers $\ket{\phi_i} = U\ket{i}$.
The wavefunction is then expanded as $\ket{\psi(\tau)}=\sum_{i}\widetilde{c_i}(\tau)\ket{\phi_i} $ where the coefficients $\widetilde{c_i}(\tau)$ evolves according to the imaginary time evolution:
\begin{equation}
\frac{\text{d} \widetilde{c_i}(\tau)}{\text{d} \tau} = -\sum_{j}({\widetilde{H_{ij}}}-S\delta_{ij})\widetilde{c_j}(\tau),
\end{equation}
where 
$
{\widetilde{H_{ij}}} = \bra{\phi_i}H\ket{\phi_j}
$ represents the Hamiltonian in the quantum walker basis and $S$ is an adjustable energy shift. Here one can start the energy shift $S$ with the VQE energy and adjust the value similarly to the classical FCIQMC setting to optimize the efficiency of the QC-FCIQMC process. 
To effectively realize the imaginary evolution, we need to implement the FCIQMC procedure to propagate the walkers. However, the key challenge arises during the spawning process, where we need to propagate walker $\ket{\phi_i}$ to $\ket{\phi_j}$ with probability $|\widetilde{H_{ij}}|\Delta \tau$. 
In conventional FCIQMC, this is feasible because there is only a polynomial number of nonzero matrix elements $H_{ji}$ for (classical) walkers $\{\ket{i}\}$, allowing straightforward enumeration of these nonzero terms for realizing walker propagation. However, with quantum walkers $\ket{\phi_i}$ and a complicated $U$, the number of non-zero $|\widetilde{H_{ji}}|$s can grow exponentially with the system size. If we follow the classical strategy, we would need to traverse through all walkers $\ket{\phi_j}$, requiring resources that scale exponentially with the system size. This exponential growth poses a significant challenge for implementing the QC-FCIQMC efficiently.

Here, we address this challenge using quantum computing by developing a novel approach to efficiently realize the spawning process. 
The key idea is to use a quantum circuit to efficiently generate samples according to the relative magnitudes of $|\widetilde{H_{ji}}|\Delta \tau$ with given $i$, regardless of whether the Hamiltonian $|\widetilde{H_{ji}}|$ is sparse or not. We summarize the main idea here and refer to Methods for a detailed description of the algorithm. 
First, we consider the squared form of $|\widetilde{H_{ji}}|$ as $|\widetilde{H_{ij}}|^2 = \langle i|U^\dag HU\Pi_jU^\dag H U|i \rangle$ with $\{\Pi_j = \ket{j}\bra{j}\}$ regarded as a projective measurement. This quantity $|\widetilde{H_{ij}}|^2$ can be regarded as the (unnormalized) distribution of measuring the state $U^\dag H U|i \rangle$ in the computational basis $\Pi_j$. 
Specifically, suppose the Hamiltonian is expanded as $H=\sum_{k}h_kP_k$ with coefficients $h_k$ and Pauli operators $P_k$, then $| \widetilde{H_{ij}}|^2 = \sum_{kk'}h_kh_{k'}p_{kk'}^i(j)$ with $p_{kk'}^i(j)=\mathrm{Re}\langle i|U^\dag P_k U\Pi_jU^\dag P_{k'} U|i \rangle$ satisfying $\sum_j |p_{kk'}^i(j)|\le 1$. For fixed $i,k,k'$, we use the quantum circuit in Fig.~\ref{FRAMEWORK}(f) to obtain samples according to $p_{kk'}^i(j)$.
Then we sum over $k,k'$ to obtain samples according to $|\widetilde{H_{ij}}|^2$.
Lastly, we apply the Bernoulli factory (the square root function)~\cite{keane1994bernoulli} to get samples according to $|\widetilde{H_{ij}}|$ and hence $|\widetilde{H_{ij}}|\Delta \tau$, which is the key for realizing the spawning process. We can also run the circuit of Fig.~\ref{FRAMEWORK}(g) to obtain the sign of $\widetilde{H_{ij}}$. By following this approach, we effectively sidestep the exponential scaling issues associated with walker propagation in the quantum walker basis and realize an efficient method for the spawning process in the context of QC-FCIQMC.

Apart from the spawning process, we can also efficiently estimate each $\widetilde{H_{ii}}$ using the same quantum circuit and hence realize the death or cloning process. The last annihilation step does not need the quantum computer. 
After implementing the imaginary time evolution (with initial walkers $\phi_0$), we can get the energy by the mixed energy evaluation 
\begin{equation}
E(\tau)=E_{0}+\sum_{i\neq0}\bra{\phi_i}H\ket{\phi_0}\frac{\textrm{sign}(i)N_i(\tau)}{N_0(\tau)},
\end{equation}
where $E_{0}=\langle \phi_0|H|\phi_0\rangle$
and $N_i(\tau)$ and $\textrm{sign}(i)$ are the number and  sign  of walker $\phi_i$, respectively. 
Suppose $\phi_0$ is obtained by running VQE, then our method effectively introduces corrections from all other $\phi_i$s by implementing QMC.  
This effect resonates with the quantum subspace expansion (QSE) method, which focuses on corrections from a specific subspace. However, QSE is limited to a small, pre-determined subspace, whereas our method considers the entire Hilbert space. We achieve this by using the quantum circuit of Fig.~\ref{FRAMEWORK}(f) to efficiently sample the most important states $\phi_j$ based on the magnitudes of $|\widetilde{H_{j0}}|^2$. A natural extension of our method is to apply QSE directly to the sampled states $\phi_j$ instead of running the full QMC process. This approach could potentially be more experimentally feasible and efficient, as it leverages the sampled states to improve accuracy without the computational requirements of full QMC. We leave the study of its effectiveness in future work.
We summarize the workflow in Fig.~\ref{FRAMEWORK}(c) and Algorithm~\ref{ALGORITHM}. We used VQE in the algorithm as an example, but it works for general VQAs.

\begin{algorithm}[H]
\caption{QC-FCIQMC}\label{ALGORITHM}
\begin{algorithmic}
\Require  Hamiltonian $H$, total evolution time $T$, time step $\Delta \tau$.
\Ensure Ground state energy estimation.

\State Run VQE to determine $U$ and hence $\{\ket{\phi_i}=U\ket{i}\}$.
\State Generate $N_0$ walkers $\ket{\phi_0}$ and let walker set $\mathcal{D}=\{0\}$.
\For{$n$ in range$(T/\Delta\tau)$}
\For{$i$ in $\mathcal{D}$}
	 \State Estimate $|H_{ji}|$ using the circuits in Fig.~\ref{FRAMEWORK}(f)~\cite{notealgorithm}. 
	 \For{$j$ with nonzero $|\widetilde{H_{ji}}|$}                                       \Comment{\textbf{Spawning step}}
            \parState{For each walker $\ket{\phi_i}$, spawn a new walker $\ket{\phi_j}$ with probability  $\Delta \tau|\widetilde{H_{ji}}|$.}
            \If{New walker $\ket{\phi_j}$ spawned}
            	\parState{Estimate $\widetilde{H_{ji}}/|\widetilde{H_{ji}}|$ using  the circuit in Fig.~~\ref{FRAMEWORK}(g)}
            	\parState{Label the new walker $\ket{\phi_j}$ with the sign of  $\ket{\phi_i}$ multiplied by  $-\widetilde{H_{ji}}/|\widetilde{H_{ji}}|$.}
				\State{Add $j$ to $\mathcal{D}$.}

            \EndIf

        \EndFor
        \State Estimate $p_i=\Delta \tau (H_{ii}-S)$
        \If{$p_i<0$}
       \Comment{\textbf{Death/cloning step}}
            \State{Clone each walker $\ket{\phi_i}$ with probability $\abs{p_i}$.}
        \Else
            \State{Kill each walker $\ket{\phi_i}$ with probability $p_i$.}
        \EndIf
        
		\For{$i$ in $\mathcal D$}\Comment{\textbf{Annihilation step}}

        \parState{Annihilate the walkers $\ket{\phi_i}$ with opposite signs.}       
         \EndFor
    
    \EndFor
    
\EndFor
\State Output the mixed energy.

\end{algorithmic}
\end{algorithm}

\begin{figure*}  \includegraphics[width=1.05\textwidth]{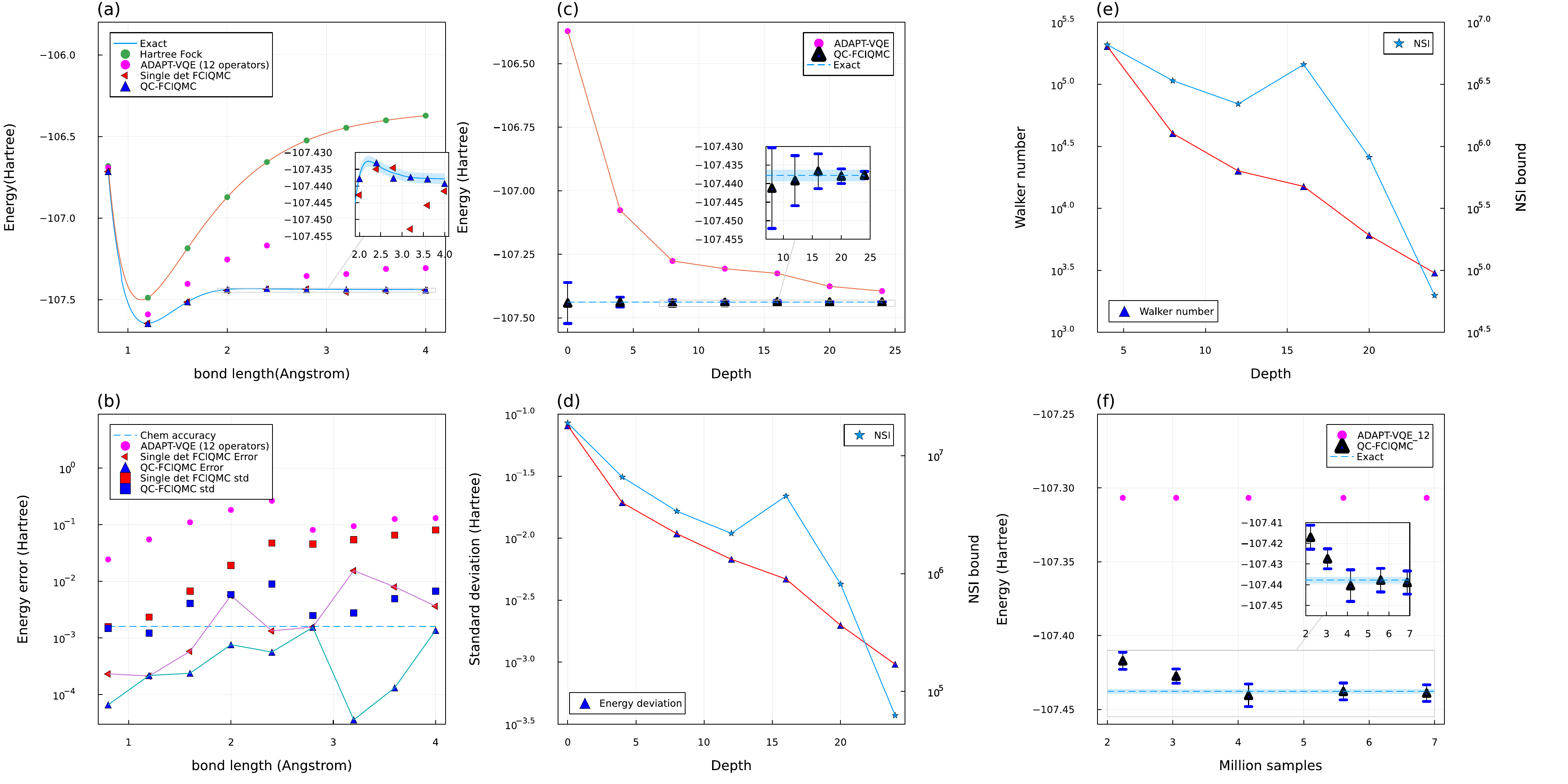}
  \caption{Numerical results of ADAPT-VQE, FCIQMC, and QC-FCIQMC for the N$_2$ molecule with 12 qubits. (a) Potential energy surface for the N$_2$ molecule with different methods under the STO-3g basis set. (b) The error and standard deviation of the average energy for different bond lengths. Here the quantum state walkers used in QC-FCIQMC are prepared with quantum circuits from ADAPT-VQE~(adding 12 fermionic operators for all bond lengths). (c) Comparison of the average energy and standard deviation for ADAPT-VQE and QC-FCIQMC with different circuit depths and a bounded number of walkers. (d) Standard deviations from (c) as well as the bounds for non-stoquastic indicator~(NSI) with $\beta=10^{-1}$.  
  (e) Log 10 plots of the numbers of walkers needed to achieve a certain precision for the energies, as well as the NSI values for reference. (f) QC-FCIQMC on N$_2$ molecule at bond length 4.0$\AA$ with sampling noise. Here we use the VQE basis generated by 12 layers of ADAPT-VQE and the $x$-axis denotes the total number of samples of the whole QC-FCIQMC process. }\label{FIGURE}
\end{figure*}

\begin{figure*}
  \includegraphics[width=0.98\textwidth]{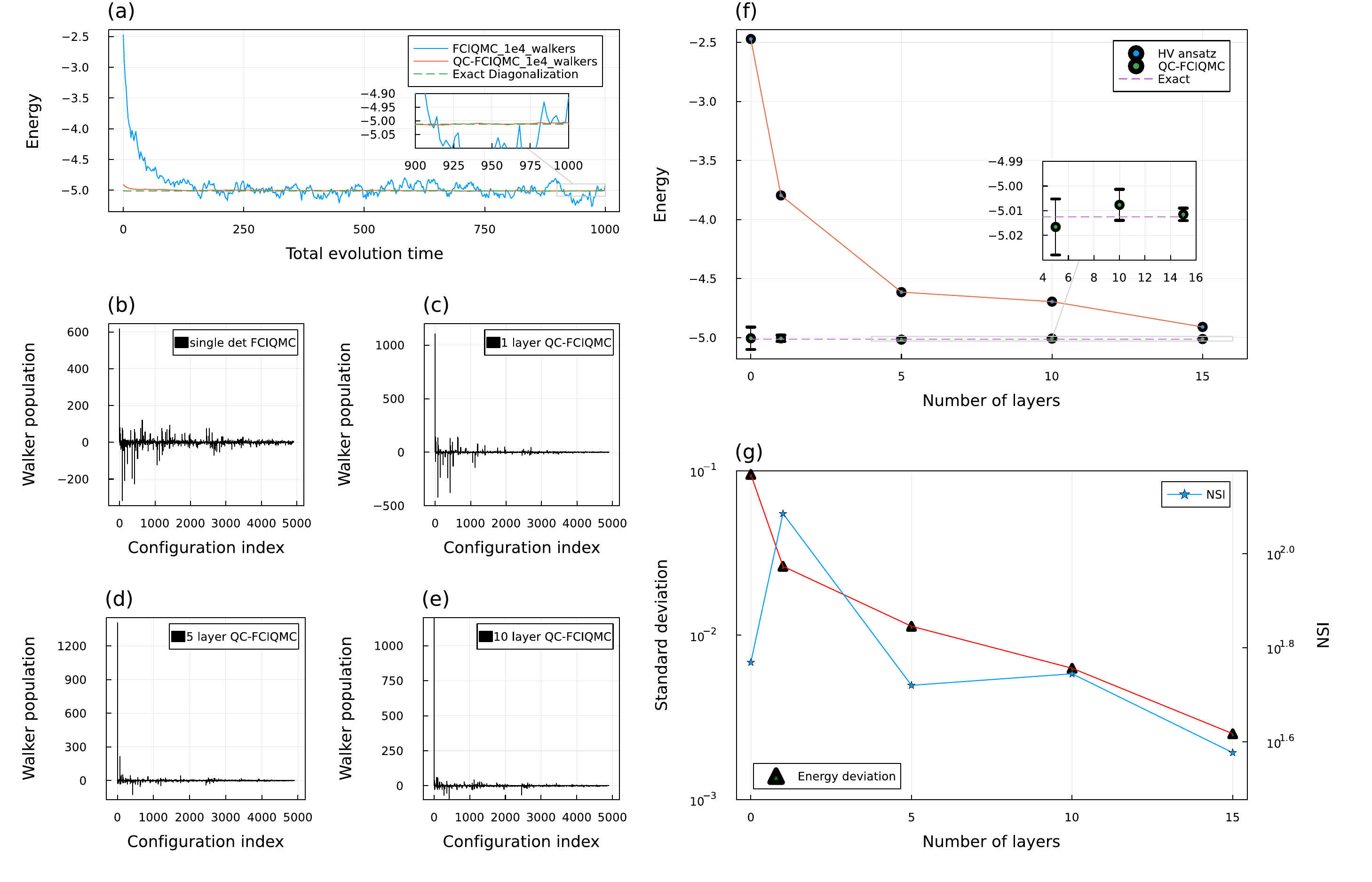}
  \caption{Numerical results of QC-FCIQMC for the 2x4 Hubbard model~($U/t=4$) with 16 qubits. The Hamiltonian variational ansatzs with 1, 5, 10, and 15 layers are implemented for comparison. The energy shift starts at $10^4$ walkers for all cases. (a) Comparison of energy fluctuation for FCIQMC and QC-FCIQMC. (b)-(e) Walker population for single determinant FCIQMC, QC-FCIQMC with HV ansatz of 1 layer, 5 layers, 10 layers respectively. (f) Energy estimation for different setups of layer number. (g) Effect of variance suppression and the NSI values.}\label{FIGURE_h}
\end{figure*}

We highlight that the sampling procedures are generated through quantum measurements, and the post-processing procedures are also efficient and independent of the system size.
The ability of quantum measurements to {generate the probability distribution of $2^n$ measurement outcomes} is crucial for efficiently performing QC-FCIQMC {without exponential measurement or computational costs.} 
This feature means QC serves not just as a solver or initial state generator but also enhances the sampling process efficiency. 
Note that the quantum circuits for estimating $| \widetilde{H_{ji}}|$ and its sign only introduce one ancillary qubit and at most double the unitary $U$ (apart from a few gates independent of $U$), adding negligible overhead compared to conventional VQE. 
Unlike conventional quantum algorithms that assume fault-tolerant quantum computers with billions of gates~\cite{kitaev1995quantum,abrams1999quantum,aspuru2005simulated,lin2022heisenberg,wan2022randomized,zeng2021universal,huo2021shallow,dong2022ground,Ge19,lu2021algorithms}, our method utilizes a shallow circuit {to effectively realize the QMC process}, suitable for the noisy intermediate-scale quantum and early fault-tolerant quantum eras. 
For a detailed discussion of the spawning process and the comparison to existing works, see Supplementary Information~\cite{notesupp}.

{We remark that our method is distinct from other recent QC-QMC proposals~\cite{Li_2019,Huggins_2022,2022arXiv220514903X,2022arXiv220509203M,2021arXiv211202190P} as well as works from classical QMC literature investigating different choices of walker basis and trial states~\cite{wouters2014projector,sandvik2010loop,clark2014stochastically}.  Compared to conventional QMC algorithms~\cite{Li_2019,2022arXiv220514903X,2022arXiv220509203M,2021arXiv211202190P}, our approach uses walkers prepared from a shallow quantum-circuit, which could be more expressive than classical basis rotations. 
Compared to the other recent QC-QMC proposals~\cite{Li_2019,Huggins_2022,2022arXiv220514903X,2022arXiv220509203M,2021arXiv211202190P}, our strategy focuses on mitigating the sign problem by replacing the walker basis instead of the trial wave function or employing other techniques such as variational quantum Monte Carlo.
Our work could also be compared to more conventional quantum eigensolver algorithms~\cite{nielsen2002quantum,albash2018adiabatic,kitaev1995quantum,abrams1999quantum,aspuru2005simulated,lin2022heisenberg,wan2022randomized,zeng2021universal,huo2021shallow,dong2022ground,Ge19,lu2021algorithms,sun2024high}, such as quantum phase estimation (we refer to Supplementary Information~\cite{notesupp} for a detailed discussion of the complexity of the methods). In summary, those quantum eigensolver algorithms generally require very deep quantum circuits (generally more than billions of gates for 100-qubit problems~\cite{sun2024high}) and hence universal quantum computers, whereas our work only needs a shallow quantum circuit and is suitable for near-term quantum computers. Compared to variational quantum eigensolver algorithms designed for near-term quantum computers, our method can achieve a much higher calculation accuracy
with shallower quantum circuits (see Sec.~\ref{Sec:numerics} for details). 

}

\section{Numerics}\label{Sec:numerics}

In this section, we present numerical experiments demonstrating the effectiveness of the QC-FCIQMC method. Our results indicate that QC-FCIQMC significantly improves the accuracy of ground state estimation using shallow quantum circuits and substantially mitigates the sign problems compared to classical FCIQMC methods.
We consider two strongly correlated systems, the $\mathrm{N_2}$ molecule 
and the Hubbard model,
and compare QC-FCIQMC to ADAPT-VQE~\cite{grimsley2019adaptive,zhang2020mutual,tang2019qubit,tang2021qubit2,fan2021circuit} and FCIQMC with single determinant walkers.

For the $\mathrm{N_2}$ molecule, we encode it using the minimum STO-3g basis set with 12 qubits by freezing the 1s and 2s orbitals.  We consider ADAPT-VQE with quantum computing, which adaptively add fermionic operators to construct the quantum circuit. In particular, we run it with varying circuit depths and observe consistently improved accuracy with increasing circuit depth. 
In Fig.~\ref{FIGURE}(a) and (b), we plot the potential energy surface and the energy difference (standard deviation), respectively, and compare the performance of different methods. Here, we set 10000 walkers to be the threshold for FCIQMC and QC-FCIQMC to start the energy shift procedure. The walker basis of QC-FCIQMC is prepared by ADAPT-VQE circuits using 12 fermionic operators. All the energy and standard deviation are evaluated by taking energies after reaching a certain total evolution time, and usually after the total number of walkers stabilizes.
We see that QC-FCIQMC is able to push the results of shallow circuits to the level of chemical accuracy across all bond lengths, much better than ADAPT-VQE and FCIQMC.
Meanwhile, the standard deviation of QC-FCIQMC is much smaller than that of FCIQMC. 
Therefore, QC-FCIQMC would require a shorter evolution time and fewer energy evaluations to obtain a precise energy estimation. 

Next, we consider how circuit depth (the number of operators added in ADAPT-VQE) affects the accuracy of our method. 
We focus on $\mathrm{N_2}$ at bond length $4.0\AA$.
We plot the effects of energy accuracy improvement and energy variance reduction with an increasing circuit depth in Fig.~\ref{FIGURE}(c). The standard deviation is reduced by two orders from a single determinant basis (FCIQMC) to ADAPT-VQE with 24 operators (QC-FCIQMC). At this depth, QC-FCIQMC achieves the (pessimistically evaluated) standard deviation within the chemical accuracy using only 10000 walkers. Meanwhile, QC-FCIQMC also visibly outperforms ADAPT-VQE for all circuit depths, making it possible to achieve high-accuracy calculations with shallow quantum circuits. 
As shown in Fig.~\ref{FIGURE}(d), the standard deviation and the NSI upper bound both decreas with increasing circuit depths. {This verifies the usefulness of the NSI and aligns well with the theoretical prediction that one can mitigate the sign problem by finding a better initial state by allowing a deeper depth of the ADAPT-VQE algorithm.}

We further consider the reduction of quantum walkers with deeper quantum circuits.
In Fig.~\ref{FIGURE} (e), we show that the total number of walkers needed to ensure a fixed level of variance exponentially decreases with the circuit depth. To reach a certain level of energy variance, QC-FCIQMC requires fewer walkers assisted with deeper ADAPT-VQE circuits. 
Lastly, we study the effect of sampling noise in quantum measurements. For each walker $i$, we 
sample a limited number of $j$s that spawns from $i$ according to the Hamiltonian magnitude. In Fig.~\ref{FIGURE} (f), we showcase the effect of sampling noise on N$_2$ with bond length 4.0$\AA$ and observe a consistent decrease of bias from the FCI result with increasing sample sizes. We find that we can obtain accurate results just using millions of samples for the entire QC-FCIQMC process. 


Next, we also test our QC-FCIQMC method on the Hubbard model.  We choose the Hubbard model of size $2\times 4$ at half-filling, which requires 16 qubits, and pick
$U/t=4$, which exhibits competition between onsite repulsion and hopping effects. We consider VQE with different layers of Hamiltonian variational ansatz~(HV ansatz)~\cite{wecker2015progress,cade2020strategies,cai2020resource} and use it in QC-FCIQMC.
 In Fig.~\ref{FIGURE_h}(a), we plot the change of the average energy with an increasing evolution time. Compared to FQCIQMC with classical walkers, our QC-FCIQMC method has much fewer energy fluctuations along the evolution, indicating the alleviation of the sign problem and the effectiveness of our method. Next, we explicitly plot the walker population with 0 (FCIQMC), 1, 5, 10 layers of the Hamiltonian variational ansatz and show the results in Fig.~\ref{FIGURE_h}(b-e).
We can clearly observe that the walker population becomes much more concentrated when using quantum state walkers with deeper quantum circuits. This observation confirms again why quantum state walkers can be effective in mitigating the sign problem.
In Fig.~\ref{FIGURE_h}(f, g), we also test the systematic energy variance reduction with an increasing number of Hamiltonian variational ansatz layers. The NSI values are also plotted. We observe the same trend as the N$_2$ case, where both NSI and the variance are exponentially reduced with an increasing number of circuit layers.

\section{Discussion \& Conclusion}

In this work, we propose a hybrid QC-QMC method that integrates quantum computing with quantum Monte Carlo to address key challenges in ground state estimation. A quantum computer is utilized to identify an improved quantum state walker basis, effectively alleviating the sign problem. Conversely, QMC enhances the expressivity of shallow quantum circuits, enabling more accurate computations with current and near-term quantum hardware. Unlike VQE, where a deep quantum circuit requires optimization and can suffer from barren plateaus and optimization difficulties, our approach uses a fixed shallow quantum circuit during the sampling process, avoiding these issues.
We acknowledge that the search for an optimal quantum circuit is a separate task within the QC-FCIQMC framework. Adaptive strategies for circuit optimization are compatible with our method and could further enhance its performance.
We derive upper bounds for non-stoquasticity indicators (NSIs), which guide the effectiveness and application of our method. The QC-FCIQMC algorithm also includes an efficient realization of the spawning process, overcoming the exponential resource requirements typically associated with this task. Benchmarking our algorithm on systems like N$_2$ and the Hubbard model demonstrates significant improvements over the use of quantum computing or FCIQMC alone.

Several promising future directions arise from this work. Firstly, the bounds we derived for non-stoquasticity indicators (NSIs) can be leveraged to explore other basis rotations as a classical approach to mitigating the sign problem. This could enhance our understanding and provide additional tools for overcoming challenges in quantum Monte Carlo simulations.
Secondly, our algorithm is designed to be compatible with current and near-term quantum hardware, and one can further study its detailed resource requirements, error mitigation strategies, and experimental realization. Additionally, integrating our circuit construction and VQE basis into deterministic selected-CI variants of FCIQMC~\cite{blunt2015semi,holmes2016heat,schriber2016communication,tubman2016deterministic} could also be explored to potentially improve performance.
Finally, we acknowledge that our method, like all approaches to the sign problem, still faces exponential scaling challenges in the worst cases. The sign problem is inherently NP-hard, as noted by \textcite{troyer2005computational}, and thus remains a formidable challenge for quantum computing. However, for practical problems, such as molecular systems with a bounded amount of static correlation or the Hubbard model, our method could significantly reduce the number of configurations needed. Recent works also show that even though QMC is not specifically designed for strongly correlated systems, it agrees well with other methods, such as AFQMC’s agreement with DMRG~\cite{qin2020absence}, highlighting the potential of QMC in solving complex problems. Together with quantum computing, we believe our approach offers a valuable pathway towards achieving quantum advantage in the near term.

\vspace{0.2cm}
\noindent\textbf{{Data Availability.---}}
The data generated for the figures in this article, including the Hamiltonians generated from VQE basis, are available at 10.5281/zenodo.13846491.

\vspace{0.2cm}

\begin{acknowledgments}
\noindent\textbf{{Acknowledgments.---}}The authors thank William Huggins and Sam McArdle for helpful discussions on the sampling procedure, Runze Chi, Tonghuan Jiang, and Ji Chen regarding the Hubbard model and QMC methods.
This work is supported by NSAF (Grant No.~U2330201), the Innovation Program for Quantum Science and Technology (Grant No. 2023ZD0300200), and the National Natural Science Foundation of China Grant (No.~12175003 and No.~12361161602). 
J.S. would like to thank the Schmidt AI in Science fellowship supported by Schmidt Futures.

\end{acknowledgments}

\bibliographystyle{achemso}
\bibliography{main}

\newpage
\appendix
\widetext

\section{Background}
In this section, we give a more detailed background introduction to variational quantum algorithms and quantum Monte Carlo. 

\subsection{Variational Quantum Eigensolver}
We focus on variational quantum eigensolver (VQE) as a typical example of general variational quantum algorithms~\cite{preskill2018quantum,PRXQuantum.2.017003,endoreview,cerezo2020variational,RevModPhys.94.015004}.
VQE is a hybrid quantum-classical method that utilizes quantum devices to prepare parameterized quantum circuits and optimize the parameters classically~\cite{peruzzo2014variational}. Through encoding the cost function, usually the corresponding Hamiltonian for the system of interest, one is able to estimate its expectation value by measuring the quantum state. 
In the following discussion, we focus on fermionic Hamiltonians. Yet, our results apply to general Hamiltonians. For a fermionic Hamiltonian, we use the Jordan-Wigner encoding ,which encodes the occupation numbers of the spin orbitals to qubits. The alternatives to JW encoding include the parity transformation and Bravyi-Kitaev transformation~\cite{bravyi2002fermionic,seeley2012bravyi}. 

Several carefully designed ansatz is proposed for the specific quantum systems concerning their unique properties. In this work, we focus on two types of algorithms: the ADAPT-VQE~\cite{grimsley2019adaptive}, and the Hamiltonian Variational Ansatz (HVA)~\cite{wecker2015progress, cade2020strategies, cai2020resource}.

ADAPT-VQE algorithms have been put forward as general protocols for adaptively growing the ansatz. The ADAPT-VQE methods build the quantum circuit iteratively by adding the operator from the operator pool with the largest gradient with respect to the loss function of the current circuit at every iteration. After adding a new operator to the circuit, one variationally optimizes the new parameter while also re-optimize all previous parameters. An iteration terminates when the norm of the gradient is smaller than a certain threshold. Several variations for this algorithm have been proposed with distinct features. The most commonly adopted operator pool is constructed according to the unitary coupled cluster ansatz with single and double excitations~(UCCSD)~\cite{romero2018strategies}. For a molecular non-relativistic Hamiltonian of the form
\begin{equation}
H=H_0 + \sum_{pq}h_{pq}a_p^{\dagger}a_q + \frac{1}{2}\sum_{pqrs}h_{pqrs}a_p^{\dagger}a_q^{\dagger}a_ra_s,   
\end{equation}
where $h_{pq}$ and $h_{pqrs}$ are coefficients obtained from one- and two-electron integrals, the coupled cluster method with single and double excitation heuristic truncated the higher excitations and gives the following ansatz based upon Hartree-Fock orbitals:
\begin{equation}
\ket{\psi_{CCSD}}=e^{T_1+T_2}\ket{\psi_{HF}}.
\end{equation}
Here $T_1$ and $T_2$ correspond to single and double excitations:
\begin{equation}
T_1 = \sum_{i\in occ,k\in virt}t_i^k a_i^\dagger a_k
\end{equation}
and
\begin{equation}
T_2 = \sum_{i>j\in occ,k>l\in virt}t_{ij}^{kl} a_i^\dagger a_j^\dagger a_k a_l
\end{equation}
The operator $e^{T}$ cannot be implemented on a quantum computer due to its non-unitarity. However, its variant $e^{T-T^\dagger}$ is unitary for a given excitation operator $T$. This is dubbed the unitary coupled cluster ansatz. Therefore, the operator pool for the ADAPT-VQE ansatz is formed by the single and double excitation operators from $T_1$ and $T_2$ but every term has to subtract its conjugate transpose. Later we will use this ansatz to construct the VQE basis for the nitrogen molecule.

The HVA ansatz is particularly suitable for the Hubbard model. The Hubbard models are celebrated for modeling strong-correlated electronic effects in many-body quantum physics. The model captures the competition between electronic hopping and on-site repulsion, which as the form
\begin{equation}\label{hubbard_h}
    H=-\sum_{i,j}t_{ij}\sum_{\sigma=\uparrow,\downarrow} a_{i\sigma}^\dagger a_{j\sigma} + U \sum_{i}n_{i\uparrow} n_{i\downarrow},
\end{equation}
where $t$ and $U$ is the hopping and repulsion strength, and $n_i$ denotes the number operator. For a square lattice, the HVA algorithm separates the Hamiltonian into three parts $H=H_v+H_h+H_U$. The former two terms are the hopping terms in Eq.~\eqref{hubbard_h} in the vertical and horizontal direction, and the last one is the on-site term. For a reference state $\ket{\psi_{\mathrm{ref}}}$, the ansatz takes the following form
\begin{equation}
    \Psi=\prod_{b=1}^B\left[\exp\left(i \frac{\theta^b_U}{2} H_U\right) \exp\left(i \frac{\theta^b_h}{2} H_h\right) \exp\left(i \frac{\theta^b_v}{2} H_v\right) \exp\left(i \frac{\theta^b_U}{2} H_U\right)\right] \ket{\psi_{\mathrm{ref}}},
\end{equation}
where $\theta$ are parameters for the ansatz, and $B$ is the total interation. With recent progress~\cite{kivlichan2018quantum}, one notices that the one-body terms in the ansatz can be compiled exactly without trotterization for they have the fermionic Gaussian form.

\subsection{Quantum Monte Carlo}

Quantum Monte Carlo~(QMC) is a class of classical computation algorithms that invokes the Monte Carlo method to approximate  properties of quantum many-body systems. Projector quantum Monte Carlo is a subclass of QMC attempting to stochastically apply  walkers to follow the imaginary-time evolution, which however generally causes the infamous sign problem in fermionic systems. The sign problem can be alleviated through different approximations, such as the fix-node approximation in Diffusion Monte Carlo~(DMC) and the phaseless approximation in auxiliary-field quantum Monte Carlo~(AFQMC). However, most techniques introduce bias while trying to ease the sign problem. In this work, we seek to alleviate the sign problem without introducing bias to the solution by combining quantum computing with a QMC algorithm. 

A hybrid algorithm of AFQMC and quantum computing has been proposed in Ref.~\cite{Huggins_2022} where they prepared a non-trivial trial state for AFQMC with the quantum device. Shadow tomography is implemented on the quantum device to calculate the overlaps between the trial state and stochastic walker states, which are the values required for energy estimation. These overlaps are easy to estimate to an additive error and are friendly for virtual correlation energy calculation without overhead on quantum devices. However, estimation of ground state energy to an additive error with the scheme in its current form requires overlap estimation to a relative error, which can be hard when some of the overlaps vanish exponentially fast as the size of the system increases.
Here in this work, we circumvent the overlap estimation by combining quantum computing with another type of projector QMC algorithm in the space of Slater determinants, called the full-configuration interaction quantum Monte Carlo~(FCIQMC) algorithm~\cite{booth2009fermion, cleland2010communications}. The FCIQMC algorithm pursues FCI-level accuracy estimation of ground-state properties with fewer resources. Now we describe its main procedure as follows.

Suppose that the wavefunction is decomposed as follows,
\begin{equation}
\ket{\psi(\tau)}=\sum_{i}c_i(\tau)\ket{i}   
\end{equation}
where $\tau$ is the imaginary time we are evolving and $\ket{i}$s are the Fock basis states, which corresponds to the single determinant configurations in the fermionic representation. Substituting this into the imaginary time Schrodinger equation
\begin{equation}
\ket{\psi(\tau)}\propto e^{-H\tau}\ket{\psi(0)},\label{ITE}
\end{equation}
taking the first-order approximation, one obtains a set of coupled differential equations for the evolution of coefficients $c_i$s:
\begin{equation}
-\frac{\text{d} c_i(\tau)}{\text{d} \tau} = \sum_{j}(H_{ij}-S\delta_{ij})c_j(\tau) \label{coeff_evolve}
\end{equation}
where $H_{ij}$s are the Hamiltonian matrix elements under a certain basis~(here the Fock basis) and $S$ is an additional energy shift on the diagonal terms. One usually starts with the Hartree-Fock energy and adjusts the value along the evolution for the purpose of walker population control. Empirical formulas are proposed in Ref.~\cite{booth2009fermion} for the adjustment of $S$ as follows
\begin{equation}
S(\tau) = S(\tau-A\Delta\tau) - \frac{\zeta}{A\Delta\tau}\text{ln}\frac{N(\tau)}{N(\tau-A\delta\tau)},
\end{equation}
where $N$ is the total walker number at a specific time, $A$ and $\zeta$ are values one can adjust according to the system of interest.

FCIQMC considers a set of walkers, who live in computational basis vector space with coefficients being $\pm 1$. The population $N_i(\tau)$ at each basis vector $\ket{i}$ is therefore an integer, obtained by the signed sum of all walkers located at this basis vector. One needs to design the QMC algorithm such that $N_i(\tau)$ is proportional to the corresponding amplitude $c_i(\tau)$ following the imaginary time dynamics. Hence, if one trotterize the time evolution into steps of size $\Delta \tau$, the FCIQMC algorithm consists of the following three steps at each time step according to eq.~(\ref{coeff_evolve}):

\begin{itemize}
 
\item Spawning: For each walker living in $\ket{i}$, spawn a child walker $\ket{j}$~($j\neq i$) with probability $\abs{H_{ij}}\Delta \tau$ up to a normalization constant. Moreover, if this quantity $H_{ij}$ is positive, then the child has the same sign as the parent, and the opposite sign otherwise.

\item Death/clone: If the quantity $H_{ii}-S>0$, then walker $\ket{\phi_i}$ dies with probability $(H_{ii}-S)\Delta \tau$; and if $H_{ii}-S<0$, then the walker clones itself with probability $\abs{(H_{ii}-S)}\Delta \tau$.

\item Annihilation: As we have alluded to, we want the signed sum of walkers to be proportional to the true amplitudes. Therefore, at the end of each time step, we take the signed sum of walkers at each basis vector and annihilate every pair of walkers with opposite signs.

\end{itemize}

\section{Sign problem in Quantum Monte Carlo} \label{sign_problem}

In this section, we give a more detailed analysis of the sign problem in QMC. We consider the non-stoquasticity indicator and  derive its upper bounds for different cases. 

\subsection{Definition and Mitigation of The Sign Problem} \label{sign-problem-defn}
The projector QMC methods are plagued by the sign problem. We briefly discuss the implications of the sign problem in the main text, and here we give a self-contained introduction and proofs of the bounds of the sign problem. As we intend to implement the imaginary time evolution (ITE) operator to an initial state, we have to normalize the resultant state due to the non-unitarity of the ITE operator. In cases of the projector QMC methods, we represent the overall wave function as a distribution of walker states over the chosen basis, and the walkers evolve upon implementing the ITE operator. The sign problem manifests when computing the expectation value of observable with the final-round walkers as we normalize the wave function in the chosen basis. To see this, let us compute the partition function of the ITE operator on the chosen basis. Here, we first consider the Fock basis for fermionic systems, and we get
\begin{equation}
\begin{aligned}
    \textup{Tr}(e^{-\beta H})&=\sum_{k=0}^{\infty}\frac{(-\beta)^k}{k!}\textup{Tr}(H^k)\\
    &= \sum_{k=0}^{\infty}\frac{(-\beta)^k}{k!}\sum_{\{\ket{\zeta}\}}\bra{\zeta_0}H\ket{\zeta_1} \bra{{\zeta_1}} H \ket{{\zeta_2}} \cdots \bra{{\zeta_{k-1}}} H \ket{{\zeta_0}}.
    \label{partition_function_path_integral}
\end{aligned}
\end{equation}
Here, $\{\ket{\zeta}\}$ is a set of complete orthogonal basis, i.e., the Fock basis in our case. Eq.~\eqref{partition_function_path_integral} gives us the path integral form of the partition function, which consists of all cyclic paths in the chosen basis. One notices that both positive and negative paths could contain in Eq.~\eqref{partition_function_path_integral}. By definition of the sign problem, the system is sign-problem-free if all the paths are positive in Eq.~\eqref{partition_function_path_integral}~\cite{troyer2005computational}. In particular, fermionic systems are known to be notoriously affected by the sign problem because the cyclic paths include cases where odd pairs of particles are exchanged, which changes the sign of the paths. The standard way to deal with the negative path is to sample according to its absolute value in Monte Carlo simulation. To quantify the sign problem, we follow the deduction of \textcite{troyer2005computational} and consider computing the expectation value of the operator $A$ with respect to the thermal state in a similar form
\begin{equation}\label{ob_path_integral}
\begin{aligned}
    \langle A\rangle
    &=\frac{\mathrm{Tr}[Ae^{-\beta H}]}{\mathrm{Tr}[e^{-\beta H}]} = \frac{\tr[Ae^{\beta G}]}{\tr[e^{\beta G}]} = \frac{\sum_{k=0}^{\infty}\frac{\beta^k}{k!}\sum_{i_0,i_1,\dots i_{k}}A_{i_0i_1}G_{i_1i_2}\cdots G_{i_{k}i_0}}{\sum_{k=0}^{\infty}\frac{\beta^k}{k!}\sum_{i_1,i_2\dots i_{k}}G_{i_1i_2}G_{i_2i_3} \cdots G_{i_{k}i_1}}\\
    &=\frac{\sum_{k=0}^{\infty}\frac{\beta^k}{k!}\sum_{i_0,i_1,\dots i_{k}}A_{i_0i_1}|G_{i_1i_2}\cdots G_{i_{k}i_0}|s_{i_1\cdots i_k,i_0}/\sum_{i_1,i_2\dots i_{k}}|G_{i_1i_2}\cdots G_{i_{k}i_0}|}{\sum_{k=0}^{\infty}\frac{\beta^k}{k!}\sum_{i_1,i_2\dots i_{k}} |G_{i_1i_2}G_{i_2i_3} \cdots G_{i_{k}i_1}|s_{i_1\cdots i_k,i_1}/\sum_{i_1,i_2\dots i_{k}}|G_{i_1i_2}\cdots G_{i_{k}i_1}|}=\frac{\langle As \rangle}{\langle s \rangle},\\
\end{aligned}
\end{equation}
where $G$ is the same as in the main text, which equals to $\alpha I-H$ with $\alpha=\textup{max}_{i} H_{ii}$, and we extract the sign of $G_{i_1i_2}G_{i_2i_3} \cdots G_{i_{k}i_1}$ as $s_{i_1\cdots i_k,i_1}$. In cases where all off-diagonal terms in the Hamiltonian are non-positive, the system is sign-problem-free as can be seen from Eq.~\eqref{ob_path_integral}, and these kinds of Hamiltonians are dubbed as ``stoquastic''~\cite{bravyi2006complexity} or ``Bosonic''. We notice that the denominator is the average sign ${\langle s \rangle}$ in Eq.~\eqref{ob_path_integral} is the ratio of the partition function and its bosonic counterpart. By definition of the partition function that is exponential of the free energy, the average sign goes down exponentially as $\mathrm{exp}(-\beta n \Delta f)$, where $\Delta f$ is the free energy difference between the original and its bosonic form system. Upon computing the expectation of an observable, the estimator is enlarged exponentially by the partition function, so as its variance
\begin{equation}
    \frac{\Delta s}{\langle s\rangle}=\frac{\sqrt{\left(\left\langle s^2\right\rangle-\langle s\rangle^2\right) / M}}{\langle s\rangle}=\frac{\sqrt{1-\langle s\rangle^2}}{\sqrt{M}\langle s\rangle} \propto \frac{e^{\beta n \Delta f}}{\sqrt{M}}.
\end{equation}
This means that the requiring number $M$ of sampled walkers needs to grow exponentially to counterbalance the variance growth. In practice, because we can only afford a finite size of samples, the sign problem often manifests itself as great statistical fluctuation in the observable estimation step. Note that it is generally hard to resolve the sign problem as \textcite{troyer2005computational} proved that the problem is $\textbf{NP}\textrm{-complete}$. In certain cases, the sign problem can be shown to be removed by a local transformation or basis rotation. However, the intrinsic hardness of solving the sign problem is characterized by their topological nature~\cite{iazzi2016topological}, and hence local transformations are not capable of curing the sign problem in the most general cases.

It is natural to think of easier ways to quantify the sign problem other than the path integral form in Eq.~\eqref{partition_function_path_integral}, and ways to mitigate the sign problem. It is also of independent interest to develop measures for the sign problem as they serve as loss functions for mitigating the sign problem~\cite{hangleiter2020easing,levy2021mitigating}, which is of great significance for extending numerical ability in classical QMC simulation of quantum many-body systems and materials. 
The criteria for the measure of the sign problem include i) faithfully reflect the severity of the sign problem for a Hamiltonian in a given basis, e.g. deviation of the sample complexity for the Hamiltonian from that of the effective bosonic form Hamiltonian; ii) general enough that it is independent of the details of implantation of the QMC algorithms; iii) efficiently computable. The non-stoquasticity, being directly related to the Hamiltonian and a computationally tractable description of the sign problem, is adopted as the measure in Ref.~\cite{hangleiter2020easing}. The authors also provide numerical evidence with randomly generated Hamiltonians to support the statement that the non-stoquasticity measures the severity of the sign problem. However, as pointed out in~Ref.~\cite{gupta2020elucidating} the off-diagonal terms do not directly induce the sign problem, but the cumulative phases of the cyclic paths in the path-integral expression do. Unfortunately, it is left unanswered to what extent the non-stoquasticity (quantitatively) contributes to the sign problem. Furthermore, it is desirable to propose a measure other than the cumulative phases, as there are exponentially many of them. In the next section, we derive rigorously the measure of the sign problem following the criteria discussed above. It is found that the upper bound for the overall effect of path-integral for the negative cyclic paths (i.e., the direct origin of the sign problem) adds up to be directly related to the negative and positive off-diagonal terms in the Hamiltonian separately. This renders our measure very much like the properties of non-stoquasticity, and we thus name the measure as non-stoquasticity indicator (NSI). Our result elucidates the relationship between non-stoquasticity and the measure of severity of the sign problem. We discuss this in detail in the next section.

To suppress the sign problem without altering the spectrum of the Hamiltonian, the general way is to implement similarity transformation with a unitary operator $U(\vec{\theta})$~\cite{hangleiter2020easing}
\begin{equation}\label{easing_sign_problem}
\begin{aligned}
    H_T(\vec{\theta})=U(\vec{\theta})^\dagger H U(\vec{\theta}).
\end{aligned}
\end{equation}
Generally speaking, an arbitrary unitary will scramble the original Hamiltonian, i.e., the Hamiltonian will become highly non-local. As a consequence, it is not possible to implement the ITE operator both in classical and quantum cases. To preserve the locality after the transformation, several proposals~\cite{marvian2019computational,hangleiter2020easing,levy2021mitigating} have been put forward. For spin systems, local Clifford and special orthogonal group~\cite{marvian2019computational,hangleiter2020easing} transformation have been considered. The former is due to the fact that the Pauli group is the normal subgroup of the Clifford group, and the latter will only scramble the observable locally while maintaining its elements real. For fermionic systems, the basis rotation transformations~\cite{levy2021mitigating} are ideal choices as they shall not alter the locality. However, both the three types of transformations are limited in their expressive power, and basically, it is $\textbf{NP}\textrm{-complete}$~\cite{marvian2019computational,hangleiter2020easing} to find the curing transformation in these settings. The above-mentioned three types of transformations are efficient-simulatable classically. To further extend the capacity of mitigating the sign problem, we suggest thinking of similarity transformation realized by quantum computing-related protocols as quantum circuits are natural building blocks to construct arbitrary unitaries. As we will discuss in the next session, we propose to alter the basis from the Fock basis to the VQE-rotated basis, which is $\{\ket{\phi_i}\equiv U(\vec{\theta})\ket{\zeta_i}\}$ with $U(\vec{\theta})$ prepared by the VQE algorithm. Let us consider the path integral form of the partition function in the VQE-rotated basis
\begin{equation}
\begin{aligned}\label{eq:nsi_upper_bound}
    &\sum_{k=0}^{\infty}\frac{(-\beta)^k}{k!}\sum_{\{\ket{\zeta}\}}\bra{\zeta_0}U(\vec{\theta})^\dagger H U(\vec{\theta})\ket{\zeta_1} \bra{{\zeta_1}} U(\vec{\theta})^\dagger H U(\vec{\theta}) \ket{{\zeta_2}} \cdots \bra{{\zeta_{k-1}}} U(\vec{\theta})^\dagger H U(\vec{\theta}) \ket{{\zeta_0}}\\
    &=\sum_{k=0}^{\infty}\frac{(-\beta)^k}{k!}\textup{Tr}(H_T(\vec{\theta})^k)=\mathrm{Tr}(e^{-\beta H_T(\vec{\theta})}),\\
\end{aligned}
\end{equation}
which matches the general way to suppress the sign problem as Eq.~\eqref{easing_sign_problem}. From this perspective, our algorithms allow an effective implementation of a rather complicated unitary to realize Eq.~\eqref{easing_sign_problem} without actually scrambling the Hamiltonian. Therefore, with the similarity transformation effectively implemented by a quantum circuit, we expect our approach to achieving better performance in easing the sign problem, and we provide quantitative measures for gauging the sign problem in the following.

\subsection{Non-Stoquasticity Indicator For Real Hamiltonians}
For any given Hamiltonian $H$, we refer to $\tilde{H}$ Bosonic form of the original Hamiltonian, which has the following form
\begin{equation}
\begin{aligned}
    \tilde{H}_{ij}&=-\left|H_{ij}\right|,\,&\textup{if}\; i\neq j;\\
    \tilde{H}_{ij}&=H_{ij},\,&\textup{otherwise}.
\end{aligned}
\end{equation}
To see the relationship between the severity of the sign problem and the non-stoquasticity of the Hamiltonian after the similarity transformation, we derive an easily computing formula by relating it to the partition function difference between $H_T$ and its Bosonic form $\tilde{H}_T$. We refer this quantity to the Non-Stoquasticity Indicator (NSI), $S(H)$ for any given Hamiltonian $H$ spanned by the basis $\{\ket{\phi}\}$, which has the form
\begin{equation}
\begin{aligned}
    S(H)&=\textup{Tr}(e^{-\beta\tilde{H}})-\textup{Tr}(e^{-\beta H})\\
    &=\textup{Tr}(e^{-\beta(\alpha I-\tilde{G})})-\textup{Tr}(e^{-\beta (\alpha I-G)})\\
    &=e^{-\beta\alpha}\bigg( \sum_{k=0}^{\infty}\frac{\beta^k}{k!} \sum_{\{\ket{\phi}\}} \big(\abs{\bra{\phi_0}G\ket{\phi_1} \bra{{\phi_1}} G \ket{{\phi_2}} \cdots \bra{{\phi_{k-1}}} G \ket{{\phi_0}}}\\
    &\quad\quad\quad\quad-\bra{\phi_0}G\ket{\phi_1} \bra{{\phi_1}} G \ket{{\phi_2}} \cdots \bra{{\phi_{k-1}}} G \ket{{\phi_0}}\big) \bigg)\\
    \label{NSI}
\end{aligned}
\end{equation}
Note that the diagonal terms in $G$ are all non-negative. We claim that the NSI is essentially the most general form that accounts for all possible negative paths in the path-integral formula for measuring the severity of the sign problem.

To arrive at the final expression for the NSI, we need to sort out all the negative path integrals in Eq.~\eqref{NSI}. We define $H_+$ and $H_-$ as
\begin{equation}
\begin{aligned}
    (H_+)_{ij}&=H_{ij},\textup{if}\, H_{ij}>0 \,~\textup{and}\, &i\neq j; (H_+)_{ij}=0,\,\textup{otherwise};\\
    (H_-)_{ij}&=H_{ij},\textup{if}\, H_{ij}<0 \,~\textup{or}\, &i=j; (H_-)_{ij}=0,\,\textup{otherwise}.
\end{aligned}
\end{equation}
According to the above definition, the corresponding $G$ will be
\begin{equation}
\begin{aligned}
    G_+&=\alpha-H_-;\\
    G_-&=\alpha-H_+.
\end{aligned}
\end{equation}
We first focus on cases where all terms in the given Hamiltonian $H$ are real. The NSI is supposed to be $-2$-fold of sum of all negative path integrals
\begin{equation}\label{NSI-real}
\begin{aligned}
    S&(H)/e^{-\beta\alpha}\\
    =&-2\sum_{k=0}^{\infty}\frac{(\beta)^{2k+1}}{(2k+1)!} \sum_{\{\eta_1,\eta_2\}} \left(G_-^{(1)}(\eta_1|\eta_2)G_+^{(2k)}(\eta_2|\eta_1) + G_-^{(3)}(\eta_1|\eta_2)G_+^{(2k-2)}(\eta_2|\eta_1) + \cdots+G_-^{(2k+1)}(\eta_1|\eta_1) \right)\\
    &-2\sum_{k=0}^{\infty}\frac{(\beta)^{2k}}{(2k)!} \sum_{\{\eta_1,\eta_2\}} \bigg(G_-^{(1)}(\eta_1|\eta_2)G_+^{(2k-1)}(\eta_2|\eta_1) + G_-^{(3)}(\eta_1|\eta_2)G_+^{(2k-3)}(\eta_2|\eta_1) + \\
    &\quad\quad\quad\quad\quad\quad\quad\quad\quad\quad\cdots+G_-^{(2k-1)}(\eta_1|\eta_2)G_+^{(1)}(\eta_2|\eta_1) \bigg),\\
\end{aligned}
\end{equation}
where $G_{-/+}^{(i)}$ stands for the path integral that contains $i$ positive/negative elements from matrix $G$, and we sum over all the possible set of paths denote by ${\eta_1,\eta_2}$ that contains the right amount of positive and negative components. The path integral form is equivalent to choosing elements from $G_-$ and $G_+$, and we have
\begin{equation}\label{NSI-real-1}
\begin{aligned}
    S&(H)/e^{-\beta\alpha}\\
    =&2\sum_{k=0}^{\infty}\frac{(\beta)^{2k+1}}{(2k+1)!}
    \left(\binom{2k+1}{1}\|G_-\|_{\mit{L}_1} \|G_+^{2k}\|_{\mit{L}_1} + \binom{2k+1}{3}\|G_-^3\|_{\mit{L}_1} \|G_+^{2k-2}\|_{\mit{L}_1} + \cdots + \binom{2k+1}{2k+1}\|G_-^{2k+1}\|_{\mit{L}_1} \right) \\
    &+2 \sum_{k=0}^{\infty}\frac{(\beta)^{2k}}{(2k)!} \left(\binom{2k}{1}\|G_-\|_{\mit{L}_1} \|G_+^{2k-1}\|_{\mit{L}_1} + \binom{2k}{3}\|G_-^3\|_{\mit{L}_1} \|G_+^{2k-3}\|_{\mit{L}_1} + \cdots + \binom{2k}{2k-1}\|G_-^{2k-1}\|_{\mit{L}_1} \|G_+\|_{\mit{L}_1} \right).\\
\end{aligned}
\end{equation}    
For $L_1$ norm of some matrix $x$, its $L_1$ norm of $m$th power is no larger than its $L_1$ norm's $m$th power for any positive integer $m$, which is $\|x^m\|_{L_1}\leq \|x\|_{L_1}^m$. We get the upper bound for the NSI,
\begin{equation}
\begin{aligned}
    S&(H)/e^{-\beta\alpha}\\
    \leq &2 \sum_{k=0}^{\infty}\frac{(\beta)^{2k+1}}{(2k+1)!} \left(\binom{2k+1}{1}\|G_-\|_{\mit{L}_1} \|G_+\|^{2k}_{\mit{L}_1} + \binom{2k+1}{3}\|G_-\|^3_{\mit{L}_1} \|G_+\|^{2k-2}_{\mit{L}_1} + \cdots + \binom{2k+1}{2k+1}\|G_+\|^{2k+1}_{\mit{L}_1} \right) \\
    &+2 \sum_{k=0}^{\infty}\frac{(-\beta)^{2k}}{(2k)!}  \left(\binom{2k}{1}\|G_-\|_{\mit{L}_1} \|G_+\|^{2k-1}_{\mit{L}_1} + \binom{2k}{3}\|G_-\|^3_{\mit{L}_1} \|G_+\|^{2k-3}_{\mit{L}_1} + \cdots + \binom{2k}{2k-1}\|G_-\|^{2k-1}_{\mit{L}_1} \|G_+\|_{\mit{L}_1} \right). \\
\end{aligned}
\end{equation}    
The above equation is just the binomial theorem with only the odd terms. By adjusting the binomial theorem to $\frac{1}{2}((b+a)^n-(b-a)^n)=\sum_{i=1,\mathrm{odd}}^n \binom{n}{i} a^x b^{n-x}$, we get
\begin{equation}
\begin{aligned}\label{NSI_1}
    S(H)/e^{-\beta\alpha}=&\sum_{k=0}^{\infty}\frac{(\beta)^{2k+1}}{(2k+1)!} 
    \left((\|G_+\|_{\mit{L}_1}+\|G_-\|_{\mit{L}_1})^{2k+1}-(\|G_+\|_{\mit{L}_1}-\|G_-\|_{\mit{L}_1})^{2k+1}\right) \\
    &+\sum_{k=0}^{\infty}\frac{(\beta)^{2k}}{(2k)!} \left((\|G_+\|_{\mit{L}_1}+\|G_-\|_{\mit{L}_1})^{2k}-(\|G_+\|_{\mit{L}_1}-\|G_-\|_{\mit{L}_1})^{2k}\right).\\
\end{aligned}
\end{equation}    
At this point, one finds that the power series in Eq.~\eqref{NSI_1} contains either odd or even terms, which fit the definitions for the hyperbolic sine and cosine functions,
\begin{equation}\label{NSI-bound}
\begin{aligned}
S(H)/e^{-\beta\alpha}&=\textup{sinh}\left(\beta(\|G_+\|_{\mit{L}_1}+\|G_-\|_{\mit{L}_1})\right)-\textup{sinh}(\beta(\|G_+\|_{\mit{L}_1}-\|G_-\|_{\mit{L}_1}))\\
&\quad+\textup{cosh}(\beta(\|G_+\|_{\mit{L}_1}+\|G_-\|_{\mit{L}_1}))-\textup{cosh}(\beta(\|G_+\|_{\mit{L}_1}-\|G_-\|_{\mit{L}_1}))\\
    &=2\textup{exp}(\beta\|G_+\|_{\mit{L}_1}) \textup{sinh}(\beta\|G_-\|_{\mit{L}_1})\\
    &=2\textup{exp}(\beta\|\alpha I-H_-\|_{\mit{L}_1})\textup{sinh}(\beta\|H_+\|_{\mit{L}_1}).
\end{aligned}
\end{equation}
The bound for the NSI formula in Eq.~\eqref{NSI-bound} indicates that the sign problem will become exponentially severe as the off-diagonal terms in the Hamiltonian increase. However, when $\|H_+\|_{\mit{L}_1}$ reduces to precisely zero, the sign problem disappears, which meets what one would expect for stoquastic Hamiltonian. Our bound for the NSI directly links the sign problem's severity with the off-diagonal terms in the Hamiltonian.

The message we learn from the NSI formula is that a similar transformation should decrease the absolute value for both positive and negative off-diagonal elements simultaneously to mitigate the sign problem.

\subsection{Non-Stoquasticity Indicator For Complex Hamiltonians}
More generally, we consider the situation where the Hamiltonian $H$ is spanned by a complex space, the analysis becomes more complicated. Note that this is the case in the following discussion about the Hubbard model as the HVA algorithms contain complex amplitudes. The NSI by definition is the difference between the partition function of the Hamiltonian $H$ and its Bosonic form Hamiltonian $\tilde{H}$. Importantly, the diagonal terms of the Hamiltonian should be real in any basis, because they correspond to the energy $E_i=\bra{\phi_i} H\ket{\phi_i}$ so that $\alpha=\mathrm{max}_i H_{ii}$ is also a real number. To keep the NSI a real measure, we apply modulus to each trace for expansion of the exponential function, which gives us
\begin{equation}\label{NSI-complex}
\begin{aligned}
    S(H)/e^{-\beta\alpha}=\sum_{k=0}^{\infty}\frac{(\beta)^k}{k!}\textup{Tr}(\tilde{G}^k)\
    -\sum_{k=0}^{\infty}\frac{(\beta)^k}{k!}\left|\textup{Tr}(G^k)\right|.\\
\end{aligned}
\end{equation}
We again expand the trace in the path integral form
\begin{equation}\label{NSI-complex-2}
\begin{aligned}
    S(H)/e^{-\beta\alpha}&= \sum_{k=0}^{\infty}\frac{(\beta)^k}{k!}\sum_{\{\ket{\phi_i},\forall i\}}\bra{\phi_0}\tilde{G}\ket{\phi_1} \bra{{\phi_1}} \tilde{G} \ket{{\phi_2}} \cdots \bra{{\phi_{k-1}}} \tilde{G} \ket{{\phi_0}}\\
    &\quad
    - \sum_{k=0}^{\infty}\frac{(\beta)^k}{k!}\left|\sum_{\{\ket{\phi_i},\forall i\}}\bra{\phi_0}G\ket{\phi_1} \bra{{\phi_1}} G \ket{{\phi_2}} \cdots \bra{{\phi_{k-1}}} G \ket{\phi_0}\right|.\\
\end{aligned}
\end{equation}
We separate the phase factor from the modulus of $H$ for each element in the matrix, $G=|G|\Modot P(G)$, where the $\Modot$ and $P(G)$ denote element-wise multiplication and the phase factor matrix for $H$, respectively. We have
\begin{equation}\label{NSI-complex-3}
\begin{aligned}
    S(H)/e^{-\beta\alpha}=& \sum_{k=0}^{\infty}\frac{(\beta)^k}{k!}\sum_{\{\ket{\phi_i},\forall i\}} \tilde{G}_{0,1}\tilde{G}_{1,2} \cdots \tilde{G}_{k-1,0} \cdot P_{0,1}(\tilde{G})P_{1,2}(\tilde{G})\cdots P_{k-1,0}(\tilde{G})\\
    &-\sum_{k=0}^{\infty}\frac{(\beta)^k}{k!}\left|\sum_{\{\ket{\phi_i},\forall i\}} |G_{0,1}||G_{1,2}| \cdots |G_{k-1,0}| \cdot P_{0,1}(G) P_{1,2}(G)\cdots P_{k-1,0}(G)\right|,\\
\end{aligned}
\end{equation}
where we denote the $i$-th row, $j$-th colomn element in matrix $X$ as $X_{i,j}$. Note that every nonzero element in $P(\tilde{G})$ equals $1$ in Eq.~\eqref{NSI-complex-3}, so we can omit this term. Note also that $\tilde{G}$ equals to $|G|$ so that
\begin{equation}\label{NSI-complex-4}
\begin{aligned}
S&(H)/e^{-\beta\alpha}\\
    =& \sum_{k=0}^{\infty}\frac{(\beta)^k}{k!}\sum_{\{\ket{\phi_i},\forall i\}} \tilde{G}_{0,1}\tilde{G}_{1,2} \cdots \tilde{G}_{k-1,0}
    -\sum_{k=0}^{\infty}\frac{(\beta)^k}{k!}\left|\sum_{\{\ket{\phi_i},\forall i\}} \tilde{G}_{0,1}\tilde{G}_{1,2} \cdots \tilde{G}_{k-1,0} \cdot P_{0,1}(G) P_{1,2}(G)\cdots P_{k-1,0}(G)\right|\\
    =&\sum_{k=0}^{\infty}\frac{(\beta)^k}{k!}\left(\sum_{\{\ket{\phi_i},\forall i\}} \tilde{G}_{0,1}\tilde{G}_{1,2} \cdots \tilde{G}_{k-1,0}
    -\left|\sum_{\{\ket{\phi_i},\forall i\}} \tilde{G}_{0,1}\tilde{G}_{1,2} \cdots \tilde{G}_{k-1,0} \cdot P_{0,1}(G) P_{1,2}(G)\cdots P_{k-1,0}(G)\right|\right).\\
\end{aligned}
\end{equation}
Notably, the last summation in Eq.~\eqref{NSI-complex-4} involves complex terms that would cancel each out. We loose this condition and attain the upper bound
\begin{equation}\label{NSI-complex-5}
\begin{aligned}
    S(H)/e^{-\beta\alpha}\leq&\sum_{k=0}^{\infty}\frac{(\beta)^k}{k!}\sum_{\{\ket{\phi_i},\forall i\}} \tilde{G}_{0,1}\tilde{G}_{1,2} \cdots \tilde{G}_{k-1,0} \left|1- P_{0,1}(G) P_{1,2}(G)\cdots P_{k-1,0}(G)\right|.
\end{aligned}
\end{equation}
In the next step, we relax the bound by decoupling the product of $\tilde{G}_{i,j}$ from the modulus term. Also, because all non-zero terms in $\tilde{G}$ are positive, we also apply bounds for its products of elements following the same deduction as Eq.~\eqref{NSI-real-1}. We have
\begin{equation}\label{NSI-complex-6}
\begin{aligned}
    S(H)/e^{-\beta\alpha}\leq&\sum_{k=0}^{\infty}\frac{(\beta)^k}{k!}\|G\|_{\mit{L}_1}^k\sum_{\{\ket{\phi_i},\forall i\}} \left|1- P_{0,1}(G) P_{1,2}(G)\cdots P_{k-1,0}(G)\right|.
\end{aligned}
\end{equation}
Now, let us consider two arbitrary phase factors $p_i=e^{i\theta_i}$ and $p_j=e^{i\theta_j}$. We can establish the triangle inequality $|1-p_i p_j| \leq |1-p_i|+|1-p_j|$. By applying the inequality iteratively to the last modulus part of Eq.~\eqref{NSI-complex-6}, we extract one phase factor out each time from the product. The resultant formula is
\begin{equation}\label{NSI-complex-7}
\begin{aligned}
    S(H)/e^{-\beta\alpha}&\leq \sum_{k=0}^{\infty}\frac{(\beta)^k}{k!}\|G\|_{\mit{L}_1}^k\sum_{\{\ket{\phi_i},\forall i\}} \left(\sum_{r=0}^{k-1}| 1-P_{r,r+1}(G)|\right)\\
    &= \sum_{k=0}^{\infty}\frac{(\beta)^k}{k!}\|G\|_{\mit{L}_1}^k\sum_{r=0}^{k-1} \left(\sum_{\{\ket{\phi_i},\forall i\}}| 1-P_{r,r+1}(G)|\right),
\end{aligned}
\end{equation}
where we have implicitly prescribed that the next term for $k-1$ is $0$. In the second line of Eq~\eqref{NSI-complex-7}, we also exchange the last two summations. By doing so, one finds that the summation runs over all possible elements for two independent basis sets $\{\ket{\phi_r}\}$ and $\{\ket{\phi_{r+1}}\}$. Then, it is obvious that the outside summation will get $k$ equal results. We get
\begin{equation}\label{NSI-complex-8}
\begin{aligned}
    S(H)/e^{-\beta\alpha}&= \sum_{k=0}^{\infty}\frac{(\beta)^k}{k!}\|G\|_{\mit{L}_1}^k \cdot k\left(\sum_{r,l}| 1-P_{r,l}(G)|\right).\\
\end{aligned}
\end{equation}
Because the two basis sets index $\{\ket{\phi_r}\}$ and $\{\ket{\phi_{r+1}}\}$ are independent, we change the indices to $r$ and $l$. We note that all non-zero elements in the $\tilde{G}$ matrix have phase factor $1$. Thus, we change the element-wise subtraction in Eq.~\eqref{NSI-complex-8} to matrix-wise subtraction in the form of the phase matrix $P$ and obtain
\begin{equation}\label{NSI-complex-9}
\begin{aligned}
    S(H)/e^{-\beta\alpha}&= \beta\|G\|_{\mit{L}_1}\|P(\tilde{G})-P(G)\|_{\mit{L}_1} \sum_{k=0}^{\infty}\frac{(\beta)^k}{k!} \|G\|_{\mit{L}_1}^k\\
    &= \beta\|G\|_{\mit{L}_1}\|P(\tilde{G})-P(G)\|_{\mit{L}_1}\textup{exp}(\beta \|G\|_{\mit{L}_1}).
\end{aligned}
\end{equation}
It is worth noting that Eq.~\eqref{NSI-complex-9} states that the sign problem relates to the norm of the Hamiltonian and also its phases of off-diagonal terms. As all phase factors of elements in $G$ approximate $1$, the sign problem alleviates. Equivalently, the more the Hamiltonian $H$ is closer to its stoquastic counterpart $\tilde{H}$, the more the sign problem is relieved.


\subsection{Initial-State-Related Non-Stoquasticity Indicator}
For our setting, it is particularly intriguing to investigate to what extent the quality of the initial state influences the sign problem. By quality, we mean the fidelity of the initial state compared to the ground state. For projector QMC algorithms, as we sum up the walkers at the final step, the overall stochastic implementation of the ITE process is equivalent to the following formula,
\begin{equation}
\begin{aligned}
    \frac{e^{-\beta H}\ket{\zeta_0}}{\bra{\zeta_0}e^{-\beta H}\ket{\zeta_0}}.
    \label{ITE-fidelity}
\end{aligned}
\end{equation}
Here, $\ket{\zeta_i}$ is the $i$-th Fock state, and the $\ket{\zeta_0}$ is the Hartree Fock state, which is usually chosen to be the initial state for various kinds of projector QMC algorithms~\cite{zhang1997constrained,booth2009fermion}.

Evidently, the partition function in Eq.~\eqref{ITE-fidelity} is related to the initial state and the basis we choose. Following the above initial-state-agnostic analysis for the partition function, here we discuss bounds for NSI equation related to the initial state. For our cases, we consider the setting that the initial state to be the VQE-optimized state $\ket{\psi_0}=U(\vec{\theta})\ket{\zeta_0}$. That is we are interested in the phenomenon that when the VQE-prepared state approaches the ground state, how does the NSI behave? The main difference of the partition function from the above analysis is that the begins and ends of path integrals are restricted to the initial state $\ket{\psi_0}$. When the expression of Hamiltonian $H$ is real in the basis $\{\phi_i\}$, the deduction of the NSI is similar to Eq.~\eqref{NSI-real}. We have

\begin{equation}\label{init-NSI-real}
\begin{aligned}
    S&(H,\psi_0)/e^{-\beta\alpha}=\frac{\bra{\psi_0}e^{-\beta \tilde{H}}\ket{\psi_0}-\bra{\psi_0}e^{-\beta H}\ket{\psi_0}}{e^{-\beta\alpha}}\\
    =&\sum_{k=0}^{\infty}\frac{\beta^k}{k!} \sum_{\{\ket{\phi}\}} \left(\abs{\bra{\psi_0}G\ket{\phi_1} \bra{{\phi_1}} G \ket{{\phi_2}} \cdots \bra{{\phi_{k-1}}} G \ket{{\psi_0}}}
    -\bra{\psi_0}G\ket{\phi_1} \bra{{\phi_1}} G \ket{{\phi_2}} \cdots \bra{{\phi_{k-1}}} G \ket{{\psi_0}}\right)
\end{aligned}
\end{equation}
It is not too hard to conceive that as the fidelity of the VQE-prepared state goes up, the off-diagonal terms in the $0$th row of matrix $G$, $G(0)$, will be suppressed, and become zeros when it reaches the ground state. To study the relationship between the fidelity $|\psi_0\rangle$ and bound for the NSI, we consider the case that the second line in Eq.~\eqref{init-NSI-real} contains $(l-1)$-degree of $G(0,0)$ from the beginning, and $(r-1)$-degree of $G(0,0)$ from the end of the path integral. Here, we denote $G(0,0)$ as the $0$-th row and column term in $G$. We sum over all possible $l$ and $r$. There are two cases where negative paths involve, which are: i) a negative intermediate path both ends either both positive or negative; ii) a positive intermediate path with one end positive and the other negative. This gives us
\begin{equation}\label{init-NSI-real-1}
\begin{aligned}
S&(H,\psi_0)/e^{-\beta\alpha}\\
    \leq& 2\sum_{k=0}^{\infty}\frac{(\beta)^{k}}{k!}\Biggl(
    \sum_{0\leq l< r \leq k-1} G(0,0)^{l-1+k-r}\biggl(\sum_{w,x,y,z=1}^{k-1}(G(0,w)_+ \cdot G(0,x)_+ + G(0,y)_- \cdot G(0,z)_-)\\
    &\cdot\sum_{\{\eta_1,\eta_2\}}
    \left|G_-^{(1)}(\eta_1|\eta_2)G_+^{(r-l-2)}(\eta_2|\eta_1) + G_-^{(3)}(\eta_1|\eta_2)G_+^{(r-l-4)}(\eta_2|\eta_1) + \cdots \right|\\
    &+\binom{2}{1}\sum_{x,y=1}^{n-1}G(0,x)_+\cdot |G(0,y)_-|\cdot\sum_{\{\eta_1,\eta_2\}} (G_+^{(r-l-1)}(\eta_1|\eta_1) + G_+^{(r-l-3)}(\eta_1|\eta_2)G_-^{(2)}(\eta_2|\eta_1) +\cdots )\biggl)
    +G(0,0)^{k-1}\sum_{x=1}^{n-1}|G(0,x)_-| \Biggl),\\
\end{aligned}
\end{equation}
where we denote the $i$-th row, $j$-th column element of $G$ as $G(i,j)$. With the same claim as Eq.~\eqref{NSI-real-1}, we have
\begin{equation}\label{init-NSI-real-2}
\begin{aligned}
S&(H,\psi_0)/e^{-\beta\alpha}\\
    \leq& \sum_{k=0}^{\infty}\frac{(\beta)^{k}}{k!}\Biggl(
    \sum_{0\leq l< r \leq k-1}G(0,0)^{l-1+k-r}\Bigl((\|G(0,\setminus0)_+\|_{\mit{L}_1} \cdot \|G(0,0)_+\|_{\mit{L}_1} + \|G(0,\setminus0)_-\|_{\mit{L}_1} \cdot \|G(0,\setminus0)_-\|_{\mit{L}_1}) \\
    &\cdot \left( (\|G_+\|_{\mit{L}_1}+\|G_-\|_{\mit{L}_1})^{r-l-1} - (\|G_+\|_{\mit{L}_1}-\|G_-\|_{\mit{L}_1})^{r-l-1}\right) \\
    &+2\|G(0,\setminus0)_+\|_{\mit{L}_1}\cdot \|G(0,\setminus0)_-\|_{\mit{L}_1}
    \cdot \left( (\|G_+\|_{\mit{L}_1}+\|G_-\|_{\mit{L}_1})^{r-l-1} + (\|G_+\|_{\mit{L}_1}-\|G_-\|_{\mit{L}_1})^{r-l-1}\right)\Bigl)\\
    &+2\binom{k-1}{1}G(0,0)^{k-1}\|G(0,\setminus0)_-\|_{\mit{L}_1}\Biggl).\\
\end{aligned}
\end{equation}
Here, $G(i,\setminus j)$ denotes the set of elements of $G$ in row $i$ and columns other than $j$. To proceed, we rearrange the terms in Eq.~\eqref{init-NSI-real-2}, and complete two squares
\begin{equation}\label{init-NSI-real-3}
\begin{aligned}
    S&(H,\psi_0)/e^{-\beta\alpha}\\
    =&\sum_{k=0}^{\infty}\frac{(\beta)^{k}}{k!}\Biggl(
    \sum_{0\leq l< r \leq k-1}G(0,0)^{l-1+k-r}\Bigl((\|G(0,\setminus0)_+\|_{\mit{L}_1}+\|G(0,\setminus0)_-\|_{\mit{L}_1})^2\cdot (\|G_+\|_{\mit{L}_1}+\|G_-\|_{\mit{L}_1})^{r-l-1}\\
    &-(\|G(0,\setminus0)_+\|_{\mit{L}_1}-\|G(0,\setminus0)_-\|_{\mit{L}_1})^2\cdot(\|G_+\|_{\mit{L}_1}-\|G_-\|_{\mit{L}_1})^{r-l-1}\Bigl)
    +2(k-1)G(0,0)^{k-1}\|G(0,\setminus0)_-\|_{\mit{L}_1}\Biggl).
\end{aligned}
\end{equation}
Next, we substitute $r-l$ to $\Delta$, and Eq.~\eqref{init-NSI-real-3} becomes
\begin{equation}\label{init-NSI-real-4}
\begin{aligned}
    S&(H,\psi_0)/e^{-\beta\alpha}\\
    =&\sum_{k=0}^{\infty}\frac{(\beta)^{k}}{k!}\Biggl(
    \sum_{1\leq \Delta \leq k-1}G(0,0)^{k-1-\Delta}\biggl(\binom{k-\Delta}{1}\cdot(\|G(0,\setminus0)_+\|_{\mit{L}_1}+\|G(0,\setminus0)_-\|_{\mit{L}_1})^2\cdot (\|G_+\|_{\mit{L}_1}+\|G_-\|_{\mit{L}_1})^{\Delta-1}\\
    &-\binom{k-\Delta}{1}\cdot(\|G(0,\setminus0)_+\|_{\mit{L}_1}-\|G(0,\setminus0)_-\|_{\mit{L}_1})^2\cdot(\|G_+\|_{\mit{L}_1}-\|G_-\|_{\mit{L}_1})^{\Delta-1}\biggl)\\
    &
    +2(k-1)G(0,0)^{k-1}\|G(0,\setminus0)_-\|_{\mit{L}_1}\Biggl).
\end{aligned}
\end{equation}
We use the gradient trick to the $G(0,0)$ terms in Eq.~\eqref{init-NSI-real-4} to get rid of the binomial coefficient $\binom{k-\Delta}{1}$, and this gives us
\begin{equation}\label{init-NSI-real-5}
\begin{aligned}
S&(H,\psi_0)/e^{-\beta\alpha}\\
    =&\sum_{k=0}^{\infty}\frac{(\beta)^{k}}{k!}\Biggl(
    \sum_{1\leq \Delta \leq k-1}\frac{\mathrm{d} G(0,0)^{k-\Delta} }{\mathrm{d} G(0,0)}\Bigl((\|G(0,\setminus0)_+\|_{\mit{L}_1}+\|G(0,\setminus0)_-\|_{\mit{L}_1})^2\cdot (\|G_+\|_{\mit{L}_1}+\|G_-\|_{\mit{L}_1})^{\Delta-1}\\
    &-(\|G(0,\setminus0)_+\|_{\mit{L}_1}-\|G(0,\setminus0)_-\|_{\mit{L}_1})^2\cdot(\|G_+\|_{\mit{L}_1}-\|G_-\|_{\mit{L}_1})^{\Delta-1}\Bigl)\Biggl)\\
    &+2\beta\|G(0,\setminus0)_-\|_{\mit{L}_1}\sum_{k=1}^{\infty}\frac{(\beta)^{k-1}}{(k-1)!}G(0,0)^{k-1}-\frac{2\|G(0,\setminus0)_-\|_{\mit{L}_1}}{G(0,0)}\sum_{k=0}^{\infty}\frac{(\beta)^{k}}{k!}G(0,0)^k.\\
\end{aligned}
\end{equation}
We first compute the last term in Eq.~\eqref{init-NSI-real-5}
\begin{equation}\label{init-NSI-real-6}
\begin{aligned}
S&(H,\psi_0)/e^{-\beta\alpha}\\
    =&\sum_{k=0}^{\infty}\frac{(\beta)^{k}}{k!}\Biggl(
    \frac{\mathrm{d} }{\mathrm{d} G(0,0)}
    \sum_{1\leq \Delta \leq k-1}G(0,0)^{k-\Delta}\Bigl((\|G(0,\setminus0)_+\|_{\mit{L}_1}+\|G(0,\setminus0)_-\|_{\mit{L}_1})^2\cdot (\|G_+\|_{\mit{L}_1}+\|G_-\|_{\mit{L}_1})^{\Delta-1}\\
    &-(\|G(0,\setminus0)_+\|_{\mit{L}_1}-\|G(0,\setminus0)_-\|_{\mit{L}_1})^2\cdot(\|G_+\|_{\mit{L}_1}-\|G_-\|_{\mit{L}_1})^{\Delta-1}\Bigl)\Biggl)\\
    &
    +2\|G(0,\setminus0)_-\|_{\mit{L}_1}\cdot\frac{2\beta G(0,0)-1}{G(0,0)}\mathrm{exp}(\beta G(0,0)).\\
\end{aligned}
\end{equation}
Note that the summation for $\Delta$ in Eq.~\eqref{init-NSI-real-6} is nothing but two geometric progressions. Applying the summation formula of geometric progression, we acquire
\begin{equation}\label{init-NSI-real-7}
\begin{aligned}
S&(H,\psi_0)/e^{-\beta\alpha}\\
    =&\sum_{k=0}^{\infty}\frac{(\beta)^{k}}{k!}
    \frac{\mathrm{d} }{\mathrm{d} G(0,0)}
    \Biggl( (\|G(0,\setminus0)_+\|_{\mit{L}_1}+\|G(0,\setminus0)_-\|_{\mit{L}_1})^2\cdot\frac{G(0,0)^k-G(0,0)(\|G_+\|_{\mit{L}_1}+\|G_-\|_{\mit{L}_1})^{k-1}}{G(0,0)-(\|G_+\|_{\mit{L}_1}+\|G_-\|_{\mit{L}_1})}\\
    &+(\|G(0,\setminus0)_+\|_{\mit{L}_1}-\|G(0,\setminus0)_-\|_{\mit{L}_1})^2\cdot\frac{G(0,0)^k-G(0,0)(\|G_+\|_{\mit{L}_1}-\|G_-\|_{\mit{L}_1})^{k-1}}{G(0,0)-(\|G_+\|_{\mit{L}_1}-\|G_-\|_{\mit{L}_1})}\Biggl)\\
    &+2\|G(0,\setminus0)_-\|_{\mit{L}_1}\cdot\frac{2\beta G(0,0)-1}{G(0,0)}\mathrm{exp}(\beta G(0,0))\\
\end{aligned}
\end{equation}
The next step is to compute the gradient, and we have
{\footnotesize
\begin{equation}\label{init-NSI-real-8}
\begin{aligned}
S&(H,\psi_0)/e^{-\beta\alpha}\\
    =&\sum_{k=0}^{\infty}\frac{(\beta)^{k}}{k!}
    \frac{(\|G(0,\setminus0)_+\|_{\mit{L}_1}+\|G(0,\setminus0)_-\|_{\mit{L}_1})^2}{G(0,0)-(\|G_+\|_{\mit{L}_1}+\|G_-\|_{\mit{L}_1})}\cdot\left((\|G_+\|_{\mit{L}_1}+\|G_-\|_{\mit{L}_1})^k+\left(k-1-\frac{k(\|G_+\|_{\mit{L}_1}+\|G_-\|_{\mit{L}_1})}{G(0,0)}\right)G(0,0)^k\right)\\
    &-\frac{(\|G(0,\setminus0)_+\|_{\mit{L}_1}-\|G(0,\setminus0)_-\|_{\mit{L}_1})^2}{G(0,0)-(\|G_+\|_{\mit{L}_1}-\|G_-\|_{\mit{L}_1})}\cdot\left((\|G_+\|_{\mit{L}_1}-\|G_-\|_{\mit{L}_1})^k+\left(k-1-\frac{k(\|G_+\|_{\mit{L}_1}-\|G_-\|_{\mit{L}_1})}{G(0,0)}\right)G(0,0)^k\right)\\
    &+2\|G(0,/0)_-\|_{\mit{L}_1}\cdot\frac{2\beta G(0,0)-1}{G(0,0)}\mathrm{exp}(\beta G(0,0)).\\
\end{aligned}
\end{equation}
}
By definition of the series expansion, we finally get
\begin{equation}\label{init-NSI-real-9}
\begin{aligned}
S&(H,\psi_0)/e^{-\beta\alpha}\\
    =&\frac{(\|G(0,\setminus0)_+\|_{\mit{L}_1}+\|G(0,\setminus0)_-\|_{\mit{L}_1})^2}{(G(0,0)-(\|G_+\|_{\mit{L}_1}+\|G_-\|_{\mit{L}_1}))^2}\mathrm{exp}(\beta(\|G_+\|_{\mit{L}_1}+\|G_-\|_{\mit{L}_1}))\\
    &-\frac{(\|G(0,\setminus0)_+\|_{\mit{L}_1}-\|G(0,\setminus0)_-\|_{\mit{L}_1})^2}{(G(0,0)-(\|G_+\|_{\mit{L}_1}-\|G_-\|_{\mit{L}_1}))^2}\mathrm{exp}(\beta(\|G_+\|_{\mit{L}_1}-\|G_-\|_{\mit{L}_1}))\\
    &+\frac{(\|G(0,\setminus0)_+\|_{\mit{L}_1}+\|G(0,\setminus0)_-\|_{\mit{L}_1})^2(\beta(G(0,0)+\|G_+\|_{\mit{L}_1}+\|G_-\|_{\mit{L}_1})-2)}{(G(0,0)-(\|G_+\|_{\mit{L}_1}+\|G_-\|_{\mit{L}_1}))^2}\mathrm{exp}(\beta G(0,0)) \\
    &-\frac{(\|G(0,\setminus0)_+\|_{\mit{L}_1}+\|G(0,\setminus0)_-\|_{\mit{L}_1})^2(\beta(G(0,0)+\|G_+\|_{\mit{L}_1}-\|G_-\|_{\mit{L}_1})-2)}{(G(0,0)-(\|G_+\|_{\mit{L}_1}-\|G_-\|_{\mit{L}_1}))^2}\mathrm{exp}(\beta G(0,0)) \\
    &+2\|G(0,\setminus0)_-\|_{\mit{L}_1}\cdot\frac{2\beta G(0,0)-1}{G(0,0)}\mathrm{exp}(\beta G(0,0)).
\end{aligned}
\end{equation}
Now, it is evident from Eq.~\eqref{init-NSI-real-9} that every term in the NSI is suppressed at least linearly by the off-diagonal terms in $G(0)$. The implication is that even if we only optimize the energy as for the VQE algorithms, we nonetheless mitigate the sign problem as off-diagonal terms vanish more and more in $G(0)$ as the prepared state gets closer to the ground state.

Whereas the Hamiltonian is complex, we follow a similar definition of NSI as Eq.~\eqref{NSI-complex} for the initial state-related case
\begin{equation}\label{NSI-complex}
\begin{aligned}
    S(H,\psi_0)/e^{-\beta\alpha}=\sum_{k=0}^{\infty}\frac{(\beta)^k}{k!}\bra{\psi_0}e^{-\beta \tilde{G}}\ket{\psi_0}\
    -\sum_{k=0}^{\infty}\frac{(\beta)^k}{k!}\left|\bra{\psi_0}e^{-\beta \tilde{H}}\ket{\psi_0}\right|.\\
\end{aligned}
\end{equation}
Compared to cases where the Hamiltonians are real, the formula is different from Eq.~\eqref{init-NSI-real-1} in that the paths other than $G(0,0)$ consist of complex elements. Thus, we can simply follow the deduction in the initial-state-agnostic situation. Following the same idea as in the real case, it is natural to expect that all terms in the bound contain terms quadratic or linear to $\|G(0,\setminus 0)\|_{\mathrm{L_1}}$. 

To summarize, as we discussed in subsection \ref{sign-problem-defn}, during alternating the walker basis, we effectively rotate the Hamiltonian with $U(\vec{\theta})$. This has two-fold implications for mitigating the sign problem. First, as one suppresses the off-diagonal terms in the effective Hamiltonian $H_T(\vec{\theta})$ the sign problem is exponentially mitigated. Second, from the initial state-related analysis, we know that the sign problem is alleviated at least linearly to $\|\Pi_{\perp}H\ket{\psi_0}\|$ with $\Pi_{\perp}=I-\ket{\psi_0}\bra{\psi_0}$. This instructs us that rather than approximately diagonalizing the Hamiltonian, we can seek an easier way that we only optimize our ansatz state $\ket{\psi_0}$ towards the ground state. Compared to the diagonalization of the Hamiltonian, the latter is more near-term amenable.

\section{Quantum Circuit-based QMC}
\subsection{Motivation}
Although VQE algorithms have become standard prototypes for solving ground state properties regarding quantum many-body and molecular systems in the NISQ era. Several drawbacks hinder their performance in both practical and theoretical ways:
\begin{itemize}
 
\item Due to the presence of all sorts of noises, the VQE algorithms are limited to relatively shallow quantum circuits for successful implementation even aided with quantum error mitigation approaches. Thus, the expressiveness of VQE is restricted, so the performance could be compromised especially when dealing with strong-correlated systems.

\item For their variational nature, the VQE approaches are not guaranteed to converge to the optimal solution because they can get trapped in a local minimum easily as~\cite{anschuetz2022beyond} proved that the loss landscape is packed out with local minimums.

\item Intrinsically, the gradients of the VQE algorithms vanish exponentially as one extends the circuit depth. These types of phenomena known as ``barren plateau''~\cite{mcclean2018barren} are found ubiquitous for any highly expressive circuits that reach 2-design over the Haar measure.

\end{itemize}

From the discussion above, we know that insights are required for one to carefully choose a suitable ansatz that is both expressive enough to include the optimal solution and not too strong to bear barren plateaus. Unfortunately, no understanding has been found for systematically building reasonable VQE circuits in generic cases. Besides, for practical concerns, most error mitigation strategies are not scalable and it is hardly possible for implementing VQE experiments to go beyond the reach of classical simulation.

To facilitate research for applying quantum algorithms to study quantum many-body systems in the NISQ era, herein we suggest combining the VQE with ITE algorithms, which are well-known approaches that guarantee to produce the ground state as we go to the long-time limit. Although the operations in the ITE approaches are not unitary, classical projector QMC methods have provided the solution to implement each evolution step in a statistical way. Therefore, each step in the whole process can be made unitary. Besides, the repeated-state-preparation feature of QMC allows shallow quantum circuits to be used in the experiment. Hence, the combination of the two cutting-edge methods enables NISQ-device-friendly implementations of our Quantum Circuit-based (QC)-QMC schemes. From another perspective, we prove and numerically demonstrate in the following session that the VQE methods are found to be helpful for easing sign problem for the QC-QMC algorithms. Thus, we expect our approaches can push the limit for studying quantum many-body and chemical problems.

\subsection{Algorithm}
\label{sec:algorithm}

In this section, we describe our hybrid algorithm. Following our previous discussion on sign problem, choosing an appropriate basis helps to reduce the sign problem and single Slater determinants are rather restricted in this sense. Though being an $\mathrm{NP}$-hard problem, the sign problem may be alleviated using more complicated basis sets provided by NISQ devices. In the following, we will introduce our algorithm using shallow VQE circuits to transform computational basis states~(single determinant state after Jordan-Wigner encoding) into entangled walkers that are closer to the diagonal basis of the Hamiltonian, which will be adopted into the FCIQMC framework to reduce the sign problem.

\begin{figure}[h]
    \centering
    \includegraphics[height=4cm]{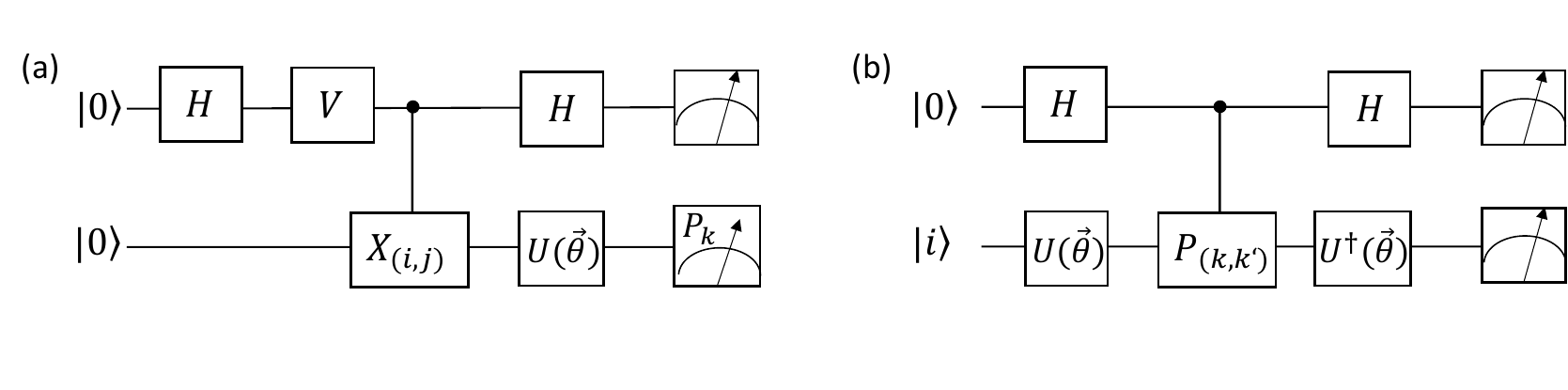}
    \caption{(a) Circuit for estimating amplitudes $\text{Re}(\bra{\phi_j}H\ket{\phi_i})$.
    One let $V=I$ when estimating the real parts and $V=S^{\dagger}=\begin{pmatrix}1 & 0\\0 & -i\end{pmatrix}$ for the imaginary parts. 
    The gate $X_{i,j}$ implements $X_i$ if the control register is $\ket{0}$ and $X_j$ if the control register is $\ket{1}$. $X_i$ implements $X$ gates on those qubits where the corresponding digits in $i$ are 1. (b) Circuit for sampling the significant amplitudes in the distribution $|\widetilde{H_{ji}}|^2$ for a fixed $i$. The gate $P_{k,k'}$ implements $P_k$ if the control register is $\ket{0}$ and $P_{k'}$ if the control register is $\ket{1}$. }
    \label{circuit_amps}
\end{figure}

Now our wavefunction is expanded in a new orthonormal basis $\{\ket{\phi_i}\}_i$:
\begin{equation}
\ket{\psi(\tau)}=\sum_{i}\widetilde{c_i}(\tau)\ket{\phi_i}   
\end{equation}
Substituting this into Eq.~(\ref{ITE}) again and taking the first-order approximation, one obtains a set of coupled differential equations for the evolution of coefficients $\widetilde{c_i}$s:
\begin{equation}
-\frac{\text{d} \widetilde{c_i}(\tau)}{\text{d} \tau} = \sum_{j}(\widetilde{H_{ij}}-S\delta_{ij})\widetilde{c_j}(\tau) \label{coeff_evolve}
\end{equation}
where $\widetilde{H_{ij}}$s are the matrix elements of $H$ under the new basis. Here we can start the energy shift $S$ with the VQE energy and adjusts the value in a similar way as in the classical FCIQMC setting. We generate the new basis by bringing the ground state of the mean-field Hamiltonian, i.e., the Hartree-Fock state, closer to the true ground state of the Hamiltonian. We achieve this by running a variational algorithm on a quantum device starting with the Hartree-Fock state and then obtaining an optimized circuit denoted as $U(\vec{\theta})$. With this circuit, we now act it on the Fock basis to form a set of new basis, $\{\ket{\phi_i}:= U(\vec{\theta})\ket{i}\}$, whose orthogonality can be easily checked as follows:
\begin{equation}
\bra{j}U(\vec{\theta})^{\dagger}U(\vec{\theta})\ket{i}=\braket{j}{i}=\delta_{ij},
\end{equation}
where $\delta$ is the Dirac delta function.
These new basis states are multi-configurational, and the optimized ground state $\{U(\vec{\theta})\ket{HF}\}$ has a bigger overlap with the exact ground state than the initial Hartree-Fock state. Given Pauli decomposition of the Hamiltonian $H=\sum_{k}h_kP_k$, the matrix elements of the Hamiltonian $\widetilde{H_{ij}}$ under basis $\{\ket{\phi_i}\}$ can be estimated on a quantum processor with the circuit in Fig.~\ref{circuit_amps} (a) for each $P_k$. With the VQE energy and the new $\widetilde{H_{ij}}$s under the new basis, we reformulate the FCIQMC procedure as follows for the sake of completeness and we summarize the whole workflow in Algorithm \ref{pseudocode}.

\begin{itemize}
 
\item Spawning: For each walker living in $\ket{\phi_i}$, spawn a child walker $\ket{\phi_j}$~($j\neq i$) with probability $\abs{\widetilde{H_{ij}}}\Delta \tau$ up to a normalization constant. Moreover, if this quantity $\widetilde{H_{ij}}$ is positive, then the child has the same sign as the parent, and the opposite sign otherwise.

\item Death/cloning: If the quantity $\widetilde{H_{ii}}-S>0$, then walker $\ket{\phi_i}$ dies with probability $(\widetilde{H_{ii}}-S)\Delta \tau$; and if $\widetilde{H_{ii}}-S<0$, then the walker clones itself with probability $\abs{(\widetilde{H_{ii}}-S)}\Delta \tau$.

\item Annihilation: At the end of each time step, we take the signed sum of walkers at each basis vector and annihilate every pair of walkers with opposite signs.

\end{itemize}

\begin{algorithm}[H]
\caption{QC-FCIQMC}\label{pseudocode}
\begin{algorithmic}
\Require  Hamiltonian $H$, total evolution time $T$, time step $\Delta \tau$.
\Ensure Ground state energy estimation.

\State Run VQE to determine $U$ and hence $\{\ket{\phi_i}=U\ket{i}\}$.
\State Generate $N_0$ walkers $\ket{\phi_0}$ and let walker set $\mathcal{D}=\{0\}$.
\For{$n$ in range$(T/\Delta\tau)$}
\For{$i$ in $\mathcal{D}$}
	 \State Estimate $|\widetilde{H_{ij}}|$ using the circuits in Fig.~\ref{circuit_amps}(b).
	 \For{$j$ with nonzero $|\widetilde{H_{ij}}|$}                                       \Comment{\textbf{Spawning step}}
            \parState{For each walker $\ket{\phi_i}$, spawn a new walker $\ket{\phi_j}$ with probability  $\Delta \tau|\widetilde{H_{ij}}|$.}
            \If{New walker $\ket{\phi_j}$ spawned}
            	\parState{Estimate $\widetilde{H_{ij}}/|\widetilde{H_{ij}}|$ using  the circuit in Fig.~~\ref{circuit_amps}(a)}
            	\parState{Label the new walker $\ket{\phi_j}$ with the sign of  $\ket{\phi_i}$ multiplied by  $-\widetilde{H_{ij}}/|\widetilde{H_{ij}}|$.}
				\State{Add $j$ to $\mathcal{D}$.}

            \EndIf

        \EndFor
        \State Estimate $p_i=\Delta \tau (\widetilde{H_{ii}}-S)$
        \If{$p_i<0$}
       \Comment{\textbf{Death/cloning step}}
            \State{Clone each walker $\ket{\phi_i}$ with probability $\abs{p_i}$.}
        \Else
            \State{Kill each walker $\ket{\phi_i}$ with probability $p_i$.}
        \EndIf
        \Comment{\textbf{Annihilation step}} 
         \parState{Annihilate the walkers $\ket{\phi_i}$ with opposite signs.}
    
    \EndFor
    
\EndFor
\State Output the mixed energy.

\end{algorithmic}
\end{algorithm}

After a certain number of evolution steps, we estimate the energy by the following formula, which is similar to the FCIQMC,
\begin{equation}
\begin{aligned}
    E(\tau)&=\frac{\bra{\phi_0}He^{-\tau H}\ket{\phi_0}}{\bra{\phi_0}e^{-\tau H}\ket{\phi_0}}\\
    &=E_{\textup{VQE}}+\sum_{i\neq0}\bra{\phi_0}H\ket{\phi_i}\frac{\phi_i(\tau)}{\phi_0(\tau)}\\
    &=E_{\textup{VQE}}+\sum_{i\neq0}\bra{\phi_0}H\ket{\phi_i}\frac{N_i(\tau)}{N_0(\tau)},
    \label{energy_estimation}
\end{aligned}
\end{equation}
where $N_i(\tau)$ is the phased (signed) number of walkers $\phi_i$ at time $\tau$. Though the number of possible nonzero $\bra{\phi_0}H\ket{\phi_i}$ can be exponentially large, the total number of computing is bounded by the number of measurements of the corresponding circuit. Besides, the computing number can be further suppressed by cutting off terms with relatively small numbers of walkers compared to $N_0$, and still control the precision within chemical accuracy. The cutoff rate is related to the fidelity of the VQE-prepared state.

One subtlety is still left for our algorithm and we propose the solution here. The classical FCIQMC $H_{ij}$ matrix under a single Slater determinant basis has $\mathcal{O}(n^2\eta^2)$ non-zero elements in any one row or column because of the Slater-Condon rules, where $n$ is the number of orbitals and $\eta$ is the number of electrons. For our VQE basis, such polynomial bound may not be guaranteed. We use the circuit in Fig.~\ref{circuit_amps} (b) to effectively sample the distribution according to $|\widetilde{H_{ij}}|^2$. 
To this end, let us re-express $|\widetilde{H_{ji}}|^2$ as
$|\widetilde{H_{ji}}|^2=\sum_{kk'}h_kh_{k'}p_{kk'}^i(j)$ with $p_{kk'}^i(j)=\mathrm{Re}\langle i|U^\dag P_k U\Pi_jU^\dag P_{k'} U|i \rangle$ satisfying $\sum_j |p_{kk'}^i(j)|\le 1$. For given $i,k,k'$, we have the normalization condition $\sum_j |p_{kk'}^i(j)|\le 1$ and we can effectively sample according to $p_{kk'}^i(j)$ using the quantum circuit.

Here, we show how to sample according to the normalized distribution $P(j|i)=|\widetilde{H_{ij}}|^2/\sum_j |\widetilde{H_{ij}}|^2$ using carefully designed quantum circuits. In particular, we focus on how to generate a new walker $j\neq i$ with probability $P(j|i)$. Since $|\widetilde{H_{ji}}|$ is hard to evaluate, we instead consider sampling according to $|\widetilde{H_{ji}}^2|$. Suppose $H=\sum_k h_k P_k$ with coefficients $h_k$ and Pauli matrices $P_k$, we have $|\widetilde{H_{ji}}|^2=\sum_{kk'}h_kh_{k'}p_{kk'}^i(j)$ with $p_{kk'}^i(j)=\mathrm{Re}\langle i|U^\dag P_k U\Pi_jU^\dag P_{k'} U|i \rangle$ satisfying $\sum_j |p_{kk'}^i(j)|\le 1$. Then we show how to sample according to $p_{kk'}^i(j)$ to effectively sample according to $|\widetilde{H_{ji}}^2|$. Our method works with the following three steps. 

\begin{enumerate}
    \item [Step 1]
 \emph{Given $i,k,k'$, generate a walker $j$ according to the unnormalized pseudo-probability distribution $p_{kk'}^i(j)=\mathrm{Re}\langle i|U^\dag P_k U\Pi_jU^\dag P_{k'} U|i \rangle$.}

    Considering the quantum circuit in Fig.~1 (f) of the main text, when the first register post-selects (i.e., measures) 0, the second register output configuration $j$ with probability $p_{k,k',+}^{i}(j) = \langle i|(U^\dag P_k+P_{k'} U)\Pi_j(U^\dag P_k+P_{k'} U)|i \rangle$. When the first register post-selects 1, the second register output bit-string $j$ with probability $p_{k,k',-}^{i}(j) = \langle i|(U^\dag P_k-P_{k'} U)\Pi_j(U^\dag P_k-P_{k'} U)|i \rangle$. One can define the following quantity thereafter
\begin{equation}
p_{k,k'}^{i}(j) = \mathrm{Re}\langle i|U^\dag P_k U\Pi_jU^\dag P_{k'} U|i \rangle  = \frac{p_{k,k',+}^{i}(j)-p_{k,k',-}^{i}(j)}{2}.
\end{equation} 
and readily check that $p_{kk'}^i(j)$ satisfies $\sum_j |p_{kk'}^i(j)|\leq 1$. Thus, $p_{kk'}^i(j)$ can be seen as a quasi-probability distribution over different $j$s (after post-processing the outcome according to the output of the first register). To generate samples according to $p_{kk'}^i(j)$, one simply gets measurement results from the quantum circuit and post-process the results. According to the outcomes of the ancillary qubit, either $\ket{0}$ or $\ket{1}$, we assign different signs $1$ or $-1$ to the outcomes, and then group the outcome with the same $\ket{j}$s together. 
For example, when we get $N_{k,k',+}^i(j)$ and $N_{k,k',-}^i(j)$ counts for outcomes $0j$ and $1j$, we equivalently get $N_{k,k'}^i(j) = N_{k,k',+}^i(j) - N_{k,k',-}^i(j)$ outcomes with respect to the quasi-probability $p_{kk'}^i(j)$. Denote the total number of samples by $N$, then $N_{k,k',+}^i(j)/N$ and  $N_{k,k',-}^i(j)/N$ are unbiased estimations of $p_{k,k',+}^{i}(j)$ and $p_{k,k',-}^{i}(j)$, respectively. Therefore, we have that $N_{k,k'}^i(j)/N$ is an unbiased estimation of the quasi-probability $p_{kk'}^i(j)$.

\item [Step 2] \emph{Given i, generate a walker $j$ according to the unnormalized probability distribution $|\widetilde{H_{ji}}|^2=\sum_{kk'}h_kh_{k'}p_{kk'}^i(j)$.}

Note that since $|\widetilde{H_{ji}}|^2$ is a linear combination of $p_{kk'}^i(j)$, we can thus obtain samples to $|\widetilde{H_{ji}}|^2$ by linearly combining samples to $p_{kk'}^i(j)$. Specifically, we first repeat the above sampling process to obtain $N_{k,k'}^i(j)$ counts for different $k,k'$. Then, we linearly combine the samples according to $|\widetilde{H_{ij}}|^2 = \sum_{kk'}h_kh_{k'}p_{kk'}^i(j)$ and obtain $N^i(j)=\sum_{kk'}h_kh_{k'}N_{k,k'}^i(j)$ counts for each $j$.  Since the samples are generated independently, we can see that the measurement outcomes effectively generate samples according to the relative strengths of $|\widetilde{H_{ij}}|^2$. We can similarly show that the normalized counts $N^i(j)/\sum_j N^i(j)$ is an unbiased estimation of $|\widetilde{H_{ij}}|^2$.  The empirical estimation of the distribution $|\widetilde{H_{ij}}|^2$ converges to its ideal value rapidly by the Bernstein bound.

\item [Step 3] \emph{Given i, generate a walker $j$ according to the unnormalized probability distribution $|\widetilde{H_{ji}}|\Delta \tau$.}

We run a Bernoulli factory~\cite{keane1994bernoulli} to apply a square root function to the probability distribution. Specifically, for any probability distribution $\vec p$, a Bernoulli factory is to generate probability distribution $f(\vec p)$. While not every function is realizable, fortunately, it is well-known that the square root function is realizable. Therefore, we can apply the square root function Bernoulli factory to the samples according to  $|\widetilde{H_{ij}}|^2$ to generate samples according to $|\widetilde{H_{ij}}|$. At last, we only need to rescale the sample counts by $\Delta \tau$ to obtain the probability distribution $|\widetilde{H_{ji}}|\Delta \tau$. We note that the whole procedure requires quantum circuit measurements of $|\widetilde{H_{ji}}|$, which also introduces shot noise and hence makes the estimator biased. However, as we show in the Appendix, we found that the effect of shot noise is negligible given a reasonable amount of sufficient samples. 
At last, to generate a walker $j$, we can randomly pick a sample from the whole set of samples.
\end{enumerate}


\subsection{Complexity Analysis}
\label{sec:complexity}
In this section, we estimate the time complexity of our algorithm. This can be divided into two parts, which are the complexity using Fig.~\ref{circuit_amps} (a) to sample the entries in the Hamiltonian matrix for estimation; and the estimation of these entries using Fig.~\ref{circuit_amps} (b). The rest is classical post-processing just like classical FCIQMC.

Assuming for a given walker $i$, we have garnered enough sample, and the walker set is denoted as $\mathcal{D}^i=\{\ket{\phi_j}\}$ using the sampling method specified in the last section. Now, let us discuss the complexity for estimation of each $\widetilde{H_{ji}}:=\langle \phi_j |H| \phi_i\rangle$ within given accuracy $\epsilon$ for all $\ket{\phi_j}$ in the walker set $\mathcal{D}^i$ using quantum circuit depicted in Figure~\ref{circuit_amps} (a). The sample complexity for achieving this goal is analyzed as follows.
\begin{lemma}[Sample complexity for estimating $\widetilde{H_{ji}}$]
Given Hamiltonian $H=\sum_{k=1}^K h_k P_k$ and target $i,j$ for $\widetilde{H_{ij}}$, let $\{\textbf{A}^{(l)}_{ijk},\textbf{B}^{(l)}_{ijk}\}_{k,l=1}^{K,L_k}$ be $M=\sum_{k=1}^K L_k$ samples that are measurement outcomes from quantum circuits given in Figure~\ref{circuit_amps} (a) with $V=I$ and $S^\dagger$, respectively, and $i,j,k$ are set to given indices. Define $\bar{\textbf{C}}_{ijk}$ as
\begin{equation}\label{eq:cijk}
    \bar{\textbf{C}}_{ijk}:=\frac{1}{L_k} \sum_{l=1}^{L_k} h_k (\textbf{A}^{(l)}_{ijk} + i \textbf{B}^{(l)}_{ijk}).
\end{equation}
Then, the estimator for $\widetilde{H_{ji}}$ defines as
\begin{equation}\label{eq:hij_estimator}
    \bar{\textbf{C}}_{ij}:=\sum_{k=1}^K \bar{\textbf{C}}_{ijk}.
\end{equation}
Also, define $h_\textrm{s}:=\sum_{k=1}^K |h_k|$.
Then for any $\epsilon>0$ and $\mu\in(0,1)$, let each $L_k$ takes the following value
\begin{equation}\label{eq:k_sample}
    L_k:=\left\lceil \frac{|h_{k}|^2\ln(4K/\mu)}{\epsilon_k^2} \right\rceil,
\end{equation}
where $\epsilon_k$ is given by
\begin{equation}\label{eq:epsilon_k}
    \epsilon_k^2:=\frac{\left|h_k\right|}{h_s} \epsilon^2.
\end{equation}
Then, when the total number of samples $M$ is given by
\begin{equation}\label{eq:total_sample}
    M:=\left\lceil \frac{h_\textrm{s}^2\ln(4K/\mu)}{\epsilon^2} \right\rceil,
\end{equation}
we have
\begin{equation}\label{eq:sample_complexity}
    \mathbb{P}\left[ |\bar{\textbf{C}}_{ij}-\widetilde{H_{ji}}|<\epsilon \right]\geq 1-\mu.
\end{equation}
\end{lemma}
\noindent \textit{Proof}. Note that Eq.~\eqref{eq:hij_estimator} serves as an unbiased estimator for evaluating the $h_k \langle \phi_j |P_k| \phi_i\rangle$, such that $\mathbb{E}[\bar{\textbf{C}}_{ijk}]\equiv h_k \langle \phi_j |P_k| \phi_i\rangle$. Therefore, Eq.~\eqref{eq:hij_estimator} serves as an unbiased estimator for estimating the $\widetilde{H_{ji}}$, and the empirical mean value approaches the expectation value when more samples are collected. For each $k\in[1,K]$, the value of $\textrm{Re}(\bar{\textbf{C}}_{ijk})$ and $\textrm{Im}(\bar{\textbf{C}}_{ijk})$ are in $[-|h_k|,|h_k|]$ as $\textbf{A}^{(l)}_{ijk}$ and $\textbf{B}^{(l)}_{ijk}$ take value in $[-1,1]$. When taking the number of samples as Eq.~\eqref{eq:k_sample} for all $k\in[1,K]$, the following formulas hold by applying Hoeffding’s inequality~\cite{hoeffding1963probability},
\begin{equation}
\left\{
\begin{aligned}
& \mathbb{P}\left[\left|\operatorname{Re}\left(\bar{\textbf{C}}_{ijk}\right)-\operatorname{Re}\left(h_k \langle \phi_j |P_k| \phi_i \rangle \right)\right|>\frac{\epsilon_k}{\sqrt{2}}\right]<\frac{\mu}{2K}, \\
& \mathbb{P}\left[\left|\operatorname{Im}\left(\bar{\textbf{C}}_{ijk}\right)-\operatorname{Im}\left(h_k \langle \phi_j |P_k| \phi_i \rangle \right)\right|>\frac{\epsilon_k}{\sqrt{2}}\right]<\frac{\mu}{2K} .
\end{aligned}
\right.
\end{equation}
By applying the triangle inequality and the union bound of the above two formulas, we have
\begin{equation}
\mathbb{P}\left[\left|\bar{\textbf{C}}_{ijk}-h_k\langle \phi_j |P_k| \phi_i \rangle\right|>\epsilon_k\right]<\frac{\mu}{K}.
\end{equation}
Finally, when choosing each $\epsilon_k$ as Eq.~\eqref{eq:epsilon_k}, the overall $\epsilon$ accuracy is achieved for evaluating $\widetilde{H_{ji}}$ by applying the union bound to all $k\in [1,K]$: \begin{equation}\label{eq:sample_complexity}
    \mathbb{P}\left[ |\bar{\textbf{C}}_{ij}-\widetilde{H_{ji}}|>\epsilon \right]< \mu.
\end{equation}
And the total number of samples $M$ is obtained by substituting Eq.~\eqref{eq:epsilon_k} into Eq.~\eqref{eq:k_sample}
\begin{equation}
\begin{aligned}
    M&=\sum_{k=1}^K L_k\\
    &=\sum_{k=1}^K\left\lceil \frac{h_s|h_{k}|\ln(4K/\mu)}{\epsilon^2} \right\rceil\\
    &=\left\lceil \frac{h_s^2\ln(4K/\mu)}{\epsilon^2} \right\rceil.
\end{aligned}
\end{equation}
Note that the sample complexity is similar to that of the energy measurement process for near-term quantum devices using the frequentist approach~\cite{wecker2015progress,romero2018strategies} as the quantity that we estimate here has a similar form to the energy.

\subsection{Comparison with Other Methods}
\label{sec:comparison}
In this section, we compare our method with recent states-of-the-art ground state energy estimation quantum algorithms~\cite{lin2020near,lin2022heisenberg,wan2022randomized,wang2022quantum,dong2022ground} in a computational complexity and resource-consumption sense in the following. 

Recently, quantum algorithms~\cite{lin2020near,lin2022heisenberg,wan2022randomized,wang2022quantum,dong2022ground} are designed regarding the ground state problems that allow a shorter circuit-depth comparing to quantum phase estimation methods and also achieve (near)-optimal scaling~\cite{lin2020near} in query complexity at the same time. Those quantum algorithms are designed for early-fault-tolerant quantum devices, where the decoherence time is supposed to be longer than near-term devices but is still limited. Besides, the number of logical qubits is limited; as such, it is favorable to employ as less as possible ancillary qubits. We provide the query complexity and maximal circuit-depth comparison between our method and state-of-the-art quantum algorithms. Particularly, we focus on three specific properties of an algorithm: i) Maximal circuit depth that needs to be implemented in one round; ii) Number of ancillary qubits needed; iii) The main subroutine that determines whether or not the method needs multi-qubit-control operations. These three properties determine whether we should implement the quantum algorithm on near-term, early-fault-tolerant, or full-fledged quantum computers. 
In addition to the advanced algorithms, we also consider two types of textbook algorithms, i.e~quantum phase estimation (QPE) method \cite{kitaev1995quantum,nielsen2002quantum} and iterative quantum phase estimation (IQPE)~\cite{whitfield2011simulation,ni2023low}. We denote the accuracy of estimating the ground state energy as $\varepsilon$, and for quantum molecular systems, a standard choice is the chemical accuracy, which is $\varepsilon_{\mathrm{chemical}}=1.6*10^{-3}$ Hartree. Also, it is typically assumed that a lower-bounded estimation $\Delta$ to the energy gap is given such that $E_1-E_0\geq\Delta$, where $E_0$ and $E_1$ is the ground- and first-excited state energy. 

Two assumptions are made for discussion of the query complexity: i) an initial state $\phi_0$ is known and its low bound $\gamma$ of overlap between the initial state and the ground state is provided, such that $|\bra{\phi_0}\Psi_0\rangle|\geq\gamma$, where $\ket{\Psi_0}$ is the ground state of $H$; ii) a lower bound $\Delta$ of the energy gap between the ground and first-excited state of the Hamiltonian is given, such that $E_1-E_0\geq\Delta$, where $E_0$ and $E_1$ are ground and first-excited state energy of $H$, respectively. We summarize the main properties of each quantum algorithm in the following table.
We summarize the main properties of each quantum algorithm in the following table.
compared to quantum phase estimation~\cite{kitaev1995quantum,nielsen2002quantum,dobvsivcek2007arbitrary} and other recently proposed ground state projection/filtering methods~\cite{lin2020near,lin2022heisenberg,wan2022randomized,wang2022quantum,dong2022ground}.
Briefly speaking, we would like to point out that our method requires much shallower quantum circuits and hence is much more resource-friendly than existing methods~\cite{kitaev1995quantum,nielsen2002quantum,dobvsivcek2007arbitrary,lin2020near,lin2022heisenberg,wan2022randomized,wang2022quantum,dong2022ground}. 
Our method is particularly tailored for NISQ hardware, whereas these existing methods~\cite{kitaev1995quantum,nielsen2002quantum,dobvsivcek2007arbitrary,lin2020near,lin2022heisenberg,wan2022randomized,wang2022quantum,dong2022ground} generally require deep quantum circuits, which generally require fault-tolerant quantum computing.
Here, we give a detailed comparison from the perspectives of theoretical complexity and resource analysis.



\begin{enumerate}
    \item \emph{Complexity analysis.---} Here, we compare the complexity (query complexity/circuit depth) of the methods. 

    \begin{itemize}
        \item We first analyze the circuit depth complexity of our QC-FCIQMC  algorithm, which effectively realizes the imaginary time evolution (ITE) process to the initial walker. Since our sampling circuit (Fig.~(f,g) in the main text) only requires one ancillary qubit and a circuit depth of $2D(U) +   {\mathcal O}(n)$, where  $D(U)$ represents the circuit depth of $U$ and $ {\mathcal O}(n)$ corresponds to the controlled Pauli operation. Since $U$ is obtained from variational quantum eigensolver, we generally have $D(U)\sim\mathcal O(n)$. Therefore, the circuit depth complexity of our method scales as $ {\mathcal O}(n)$. We note that the circuit depth does not explicitly depend on the accuracy $\varepsilon$ or the energy gap $\Delta$, because the quantum circuit is only used as the walker basis and the sampling of the walkers. Instead, the imaginary evolution time might depend on $\varepsilon$ and $\Delta$, which might thus introduce more repetitions of the quantum circuits (could be sped up with multiple quantum hardware). The feature of our method (shallower circuit depth at a cost of more circuit repetitions) exactly reflects the power of NISQ hardware, where the circuit depth is generally limited owing to non-negligible gate errors, yet we could repeatedly run the circuits. In particular, we note that our circuit-depth requirement is suitable for near-term quantum devices according to a recent experimental demonstration of VQE methods~\cite{google2020hartree,stanisic2022observing,o2022purification,guo2022scalable}.

{
\begin{table}[h]
    \centering
\caption{Quantum algorithms performance for the ground state energy estimation problem
    }
    \begin{tabular}{c|c|c|c}
      \hline\hline
        & max query/depth & \# of ancillary qubits & main subroutine \\
      \hline
       QPE~\cite{kitaev1995quantum,nielsen2002quantum} &$\mathcal{\widetilde{O}}(\varepsilon^{-1})$  & $\mathcal{O}(\mathrm{polylog}(\gamma^{-1}\varepsilon^{-1}))$ & \begin{tabular}{@{}c@{}}controlled Hamiltonian evolution/qubitization  \\ + Quantum Fourier transform\end{tabular} 
        \\
        \hline
        Iterative QPE~\cite{whitfield2011simulation,ni2023low} & $\mathcal{\widetilde{O}}(\varepsilon^{-1}\gamma^{-2})$ & $\mathcal{O}(1)$ & \begin{tabular}{@{}c@{}}controlled Hamiltonian evolution  \\ + Hadamard test\end{tabular}
        \\
        \hline
        LT~\cite{lin2022heisenberg} & $\mathcal{\widetilde{O}}(\varepsilon^{-1})$ & $\mathcal{O}(1)$ & \begin{tabular}{@{}c@{}}controlled Hamiltonian evolution  \\ + Hadamard test\end{tabular}
        \\
        \hline
        \textcite{wan2022randomized} & $\mathcal{\widetilde O}(\varepsilon^{-2}\lambda^2)$ & $\mathcal{O}(1)$ & \begin{tabular}{@{}c@{}}controlled Pauli rotation  \\ + Hadamard test\end{tabular}
        \\
        \hline
        \textcite{wang2022quantum} & $\mathcal{\widetilde{O}}(\Delta^{-1})$ & $\mathcal{O}(1)$ & \begin{tabular}{@{}c@{}}controlled Hamiltonian evolution  \\ + Hadamard test\end{tabular}
        \\
        \hline
        DLT~\cite{dong2022ground} & $\mathcal{\widetilde{O}}(\varepsilon^{-1}\gamma^{-1})$ & $\mathcal{O}(1)$ & \begin{tabular}{@{}c@{}}controlled Hamiltonian evolution  \\ + quantum signal processing\end{tabular}
        \\
        \hline
        This work & $\mathcal{O}(n)$ & $\mathcal{O}(1)$ & \begin{tabular}{@{}c@{}}controlled Pauli rotation  \\ + Hadamard test\end{tabular}

        \\
        \hline\hline
    \end{tabular}
    \label{tab:gsee-1}
\end{table}
}
        \item As for the other existing methods~\cite{lin2022heisenberg,wan2022randomized,wang2022quantum,ni2023low}, they generally need to implement the Hadamard-test quantum circuit with the controlled unitary being the  Hamiltonian evolution operator with evolution time $t$. Further operations, such as quantum Fourier transform,m might be applied, which would cause more gate depth. Here, we first focus on the maximal evolution time of these methods regarding $\varepsilon$ and $\Delta$. The complexity of the circuit depth also needs to take into account the circuits for realizing the time evolution. We summarize the maximal query of the evolution time $t$ in Table~\ref{tab:gsee-1}. For example, the work of \textcite{lin2022heisenberg} applies the maximum evolution time for the Hamiltonian evolution at a single iteration is $t=\mathcal{\widetilde{O}}(\varepsilon^{-1})$. Ref.~\cite{wang2022quantum} managed to decrease the maximal evolution time to $\mathcal{\widetilde{O}}(\Delta^{-1})$. Since $\varepsilon^{-1}$ or $\Delta^{-1}$ are also polynomial functions of $n$, the complexity of the evolution time is already generally larger than $\Omega(n)$. Furthermore, there also exists a cost for implementing $e^{-iHt}$. For example, even just considering a simple 1D Ising Hamiltonian with nearest-neighbor interaction, the gate complexity for implementing $e^{-iHt}$ is $\mathcal O(n^3)$ even for the best existing Hamiltonian method~\cite{childs2018toward}. The cost would even be larger for more complex chemistry Hamiltonians.  Therefore, the overall circuit complexity would be much larger than linear (our method). This explains why these existing methods~\cite{kitaev1995quantum,nielsen2002quantum,whitfield2011simulation,lin2022heisenberg,wan2022randomized,wang2022quantum,ni2023low} are generally considered (early) fault-tolerant algorithms~\cite{liang2023modeling}, which are beyond the capability of near-term quantum computing. 



    \end{itemize}

    \item 
    
    \emph{Resource analysis.---} Here, we also present the resource (circuit depth/gate complexity) analyses of our and other methods~\cite{lin2022heisenberg,wan2022randomized,wang2022quantum,ni2023low} for some example problems to explicitly show the advantage of our method. This comparison will not include the quantum circuits for the initial state preparation for the early fault-tolerant algorithms~\cite{lin2022heisenberg,wan2022randomized,wang2022quantum,ni2023low} and the VQE-prepared state in our method.

    From the above discussion, we know that for early fault-tolerant algorithms~\cite{lin2022heisenberg,wan2022randomized,wang2022quantum,ni2023low}, the most circuit-depth consuming part is the (controlled) Hamiltonian simulation, which is to approximate $e^{-iHt_{\textrm{max}}}$ with longest evolution time $t_{\textrm{max}}=\frac{1}{\varepsilon}$. Here, $\varepsilon$ is the accuracy in estimating the ground state energy. We will first provide analysis results for the Hamiltonian simulation and then lift the results to a controlled version.
    As such, we will compare the resources of our methods with Hamiltonian simulation methods that are suitable for early fault-tolerant devices. Note that the following resource estimation results do not include the overhead for fault tolerance (which generally requires thousands of physical qubits to encode a single logical qubit and additional operations on magic state distillation.). As suggested in the early fault-tolerant algorithms~\cite{lin2022heisenberg,dong2022ground}, the Trotter formulae are favored for the realization of the Hamiltonian simulation.
    This is because the Trotter formulae may outperform other advanced methods in gate complexity~\cite{childs2018toward,lin2022heisenberg}, despite other methods that can be more favorable for query complexity. Moreover, it is easy for the Trotter formulae to be tailored for specific quantum models, for which the 2D Fermionic Hubbard model is extensively studied~\cite{kivlichan2020improved,campbell2021early,flannigan2022propagation}, to further reduce the gate complexity. Therefore, we will consider the Hamiltonian simulation of 2D fermionic Hubbard models of $m$ fermionic modes (sites) with the Trotter formulae. The Hamiltonian is given by
    \begin{equation}
        H=-t \sum_{\langle i, j\rangle, \sigma}\left(a_{i \sigma}^{\dagger} a_{j \sigma}+a_{j \sigma}^{\dagger} a_{i \sigma}\right)+U \sum_i n_{i \uparrow} n_{i \downarrow},
    \end{equation}
    where $\sigma \in\{\uparrow, \downarrow\}$ and $n$ is the particle number operator. We will introduce the zigzag index strategy as shown in [Ref.~\cite{cade2020strategies}, FIG.~1]. That is given the 2D grid, we index each site from left to right, and then move to the next row with reverse ordering, and repeat accordingly. By Jordan-Wigner transformation (JWT), the $m$-model Hubbard model transforms to a $n=2m$-qubit system as the fermionic operators are transformed as 
    \begin{equation}
    \hat a_{j} \mapsto \frac{1}{2}\left( X_{j}+i Y_{j}\right) \bigotimes_{i=1}^{j-1} Z_{i},~ \hat a_{j}^{\dagger} \mapsto \frac{1}{2}\left( X_{j} - i Y_{j}\right) \bigotimes_{i=1}^{j-1} Z_{i}
    \end{equation}
    with Pauli operators $X$, $Y$, $Z$ acting on the $j$th qubit. Besides, we will encode spin-up and down sectors on the same site to adjacent indices.

    \begin{table}
    \begin{tabular}{l|c|r} 
    \hline
    \hline
    Gate & Gate Count & Depth \\
    \hline Single-qubit gates & $43m-40\sqrt{m}$ & 16 \\
    CNOT & $2(15 m-20 \sqrt{m}+6)$ & $6(2 \sqrt{m}+1)$\\
    \hline
    \hline
    \end{tabular}
    \caption{Adapted from Ref.~\cite{flannigan2022propagation}. Total resources (single-qubit gates including Clifford and non-Clifford $R_z$) for implementing a single time step sweep for a $m$ modes (sites) Hubbard model.}
    \label{tab:hubbard_re}
    \end{table}
    
    Next, we introduce the Hamiltonian simulation results that are specially tailored for the Hubbard model in Ref.~\cite{flannigan2022propagation}.
    It is shown in Table~\ref{tab:hubbard_re} of the three types of different gate counts for a single step of the Trotter step. Here, we consider the $1$st order Trotter formula as an instance. When considering controlled evolution, the single-qubit and CNOT gates will be transformed into CNOT plus single-qubit and Toffoli gates, respectively. Apply the well-known results that a Toffoli gate can be compiled into $6$ CNOT and $8$ single-qubit (both Clifford and non-Clifford) gates. This gives us the final result that to realize a single controlled Trotter step, the CNOT gate count is $223m-280\sqrt{m}+72$. Because the control qubit of these CNOT gates is the ancillary qubit introduced in the Hadamard test, the circuit depth is the same as the gate count. The (worst-case) single-qubit gate count is $326m-400\sqrt{m}+96$. Next, we are going to analyze the total Trotter step. As analyzed in Ref.~\cite{lin2022heisenberg,dong2022ground}, to guarantee the desired accuracy of energy estimation, we need $r$-Trotter step of a $1$st-order Trotterization for the maximal evolution time $t_{\textrm{max}}=\frac{1}{\varepsilon}$ such that 
    \begin{equation}
        r=\widetilde{\mathcal{O}}\left(\max \left(t_{\textrm{max}}C_{\text {Trotter }} \varepsilon^{-1} \gamma^{-1}, t_{\textrm{max}}\right)\right)
    \end{equation}
    where $C_{\text {Trotter }}=\mathcal{O}\left(\left(\sum_i\|H\|_i\right)^{2}\right)$, and $\gamma$ is the modulus square of the overlap between the initial state and the ground state. For a $4\times 4$ system (i.e. $m=16$), the single Trotter step will introduce $2520$ CNOT gates and $3712$ single-qubit gates. Next, we consider a rather optimistic case where $\gamma=0.25$ corresponds to an initial state of $50\%$ fidelity, $\varepsilon=10^{-1}$ and  taking $C_{\text {Trotter }}=1$. This gives us the total Trotter step $r=2.5*10^3$. Finally, the CNOT and single-qubit gate counts for this case are $6.3*10^6$ and $9.28*10^6$, respectively.

    In comparison, because our method only needs to implement controlled Pauli operations, which involve (off-diagonal) terms in the Hamiltonian, we only need to implement Pualis in the single-body terms (i.e., hopping terms) in the Hamiltonian after the JWT. As such, in the worst case corresponding to the interaction between $j$ and $j+2\sqrt{m}$-th qubits, $2\sqrt{m}=8$ CNOT gates (correspond to $8$ depth) will be introduced. By contrast, only $4$ Clifford gates (depth $2$) will be introduced and no non-Clifford gates are involved. Therefore, the maximal circuit depth for our method is $10$.
    
    From the above discussion, we can see that the high-circuit-depth overhead for advanced methods such as the RFE algorithms~\cite{lin2022heisenberg,wan2022randomized,wang2022quantum,ni2023low} demands fault tolerance naturally. Yet, our methods only require a quantum circuit of sublinear depth to the system size, which is already within reach for near-term quantum devices~\cite{google2020hartree,guo2022scalable}. Although here we only focus on the fermionic Hubbard model, our analysis can be readily extended to quantum molecular systems, where the results we believe can still be held. This is because the controlled Hamiltonian simulation for quantum molecular systems can be even more challenging. Yet, the most complicated controlled Pauli terms will be introduced by the two-body interaction terms such that a linear number of CNOT gates will be employed and the number of non-Clifford gates will stay constant.
    \end{enumerate}

\section{Numerical results}
\label{sec:numerical}
In this section, we present numerical results demonstrating the effectiveness of the QC-FCIQMC algorithm at easing the sign problem and reducing the number of walkers as well as the energy variance compared to classical FCIQMC. We pick two strongly correlated systems to ensure the robustness of our scheme, the $\mathrm{N_2}$ molecule at the dissociation limit and the $2\times 4$ Hubbard model with $U/t=4$.

\begin{figure}[h]
    \centering
    \includegraphics[height=8.0cm]{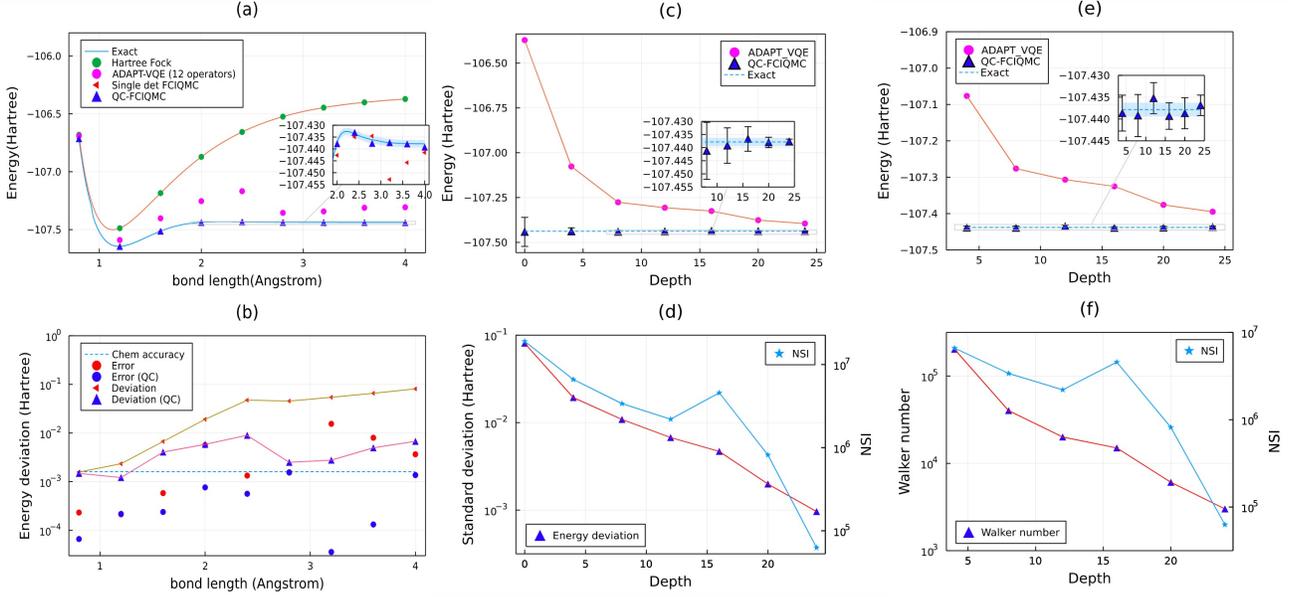}
    \caption{\scriptsize (a) Potential energy surface for the nitrogen molecule with QC-FCIQMC under the STO-3g basis set. (b) The standard deviation of energy evaluations along the QMC evolution. Here the new walker space of QC-FCIQMC is prepared with circuits from ADAPT-VQE~(adding 12 fermionic operators for all bond lengths). Even with the assistance of shallow-depth ADAPT-VQE runs, QC-FCIQMC manages to reach chemical accuracy across all bond lengths using only around 10000 walkers. We also include data from FCIQMC with single-determinant walkers for comparison. For both single determinant FCIQMC and QC-FCIQMC, we start the energy shift when the total number of walkers exceeds 10000. Under this setting, single determinant FCIQMC fails to reach chemical accuracy at several bond lengths towards the dissociation limit. Moreover, the standard deviation from the energy profiling of QC-FCIQMC is smaller than the single determinant FCIQMC for all bond lengths. In fact, one observes a bigger variance reduction with a stronger static correlation. (c) FCIQMC-assisted ADAPT-VQE results for nitrogen molecule at bond length 4.0. ADAPT-VQE energies and QC-FCIQMC energies with standard deviations for different depths of ADAPT-VQE. (d) Log plot with standard deviations from (c) as well as the non-stoquastic indicator, where choose $\beta=10^{-1}$. FCIQMC is able to push ADAPT-VQE results of different depths to chemical accuracy. Here we obtain the expected energy by taking the average of all FCIQMC energies after it converges.  We fix the total walker number at the same level by starting to implement an energy shift of $S$ when the total walker number exceeds $10^4$. The deeper the ADAPT-VQE circuit is, the smaller the energy variance FCIQMC would have. One can see with an ADAPT-VQE optimization that picks 20 operators, the energy variance obtained is already at a similar level as chemical accuracy requirement and with 24 operators the variance lies well within chemical accuracy area. (e) ADAPT-VQE energies and QC-FCIQMC energies with variances for different depths of ADAPT-VQE. (f) Log 10 plots of the respective numbers of walkers that give the energies in (e) as well as the natural log of the non-stoquastic indicator, where choose $\beta=10^{-1}$. To reach a certain level of energy variance, FCIQMC requires fewer walkers assisted with deeper ADAPT-VQE circuits. Here we choose the energy variance upper bound to be 5mHa for the number of walkers to be friendly to the running time of our code. To reach this precision, QC-FCIQMC requires about 100 times fewer walkers going from ADAPT-VQE with 4 operators to that of 24 operators. }
    \label{N2_curve}
\end{figure}

We plot the potential energy surface of $\mathrm{N_2}$ molecule with our QC-FCIQMC algorithm in Fig.~\ref{N2_curve} (a). The new walker bases for all bond lengths are prepared by ADAPT-VQE circuits that add 12 fermionic operators. In Fig.~\ref{N2_curve} (b), we plot the standard deviation from energy profile along the QC-FCIQMC evolution. We freeze the 1s and 2s orbitals and run ADAPT-VQE with 12 qubits. We also plot data from a single determinant FCIQMC for comparison. We set 10000 walkers to be the threshold for both algorithms to start the energy shift. On the one hand, we see that QC-FCIQMC is able to push the results of shallow ADAPT-VQE circuits to the level of chemical accuracy across all bond lengths, therefore easing the burden on the quantum devices. Meanwhile, using the same amount of walkers, single determinant FCIQMC fails to reach chemical accuracy at some of the bond lengths towards the dissociation limit. On the other hand, the energy deviation of QC-FCIQMC is smaller than that of single determinant FCIQMC. Therefore, QC-FCIQMC would require much fewer energy evaluations to obtain a precise energy estimation and maybe smaller total evolution time. All the energy and standard deviation are evaluated by taking energies after reaching a certain total evolution time, and usually, after the total number of walkers stabilizes. We adopt this way of evaluation for all of the following numerical experiments.

For the $\mathrm{N_2}$ molecule at bond length $4.0\AA$, we use quantum circuits from ADAPT-VQE that add different numbers of operators to generate the new basis of QC-FCIQMC. We plot the effect of energy variance reduction of QC-FCIQMC given a fixed number of walkers in Fig.~\ref{N2_curve} (c) (d). The standard deviation is reduced by close to 100 folds going from a single determinant basis to ADAPT-VQE adding 24 operators. We highlight that at this depth, QC-FCIQMC achieves a standard deviation within the range of chemical accuracy with only 10000 walkers. 

In Fig.~\ref{N2_curve} (e) (f), we conversely plot the total walker number required for QC-FCIQMC runs assisted by ADAPT-VQE circuits of different depths to maintain a similar level of precision. The total number of walkers decreases by two orders of magnitude going from the basis generated by ADAPT-VQE adding 4 operators to that generated by ADAPT-VQE adding 24 operators.

\begin{figure}[h]
    \centering
    \includegraphics[height=6cm]{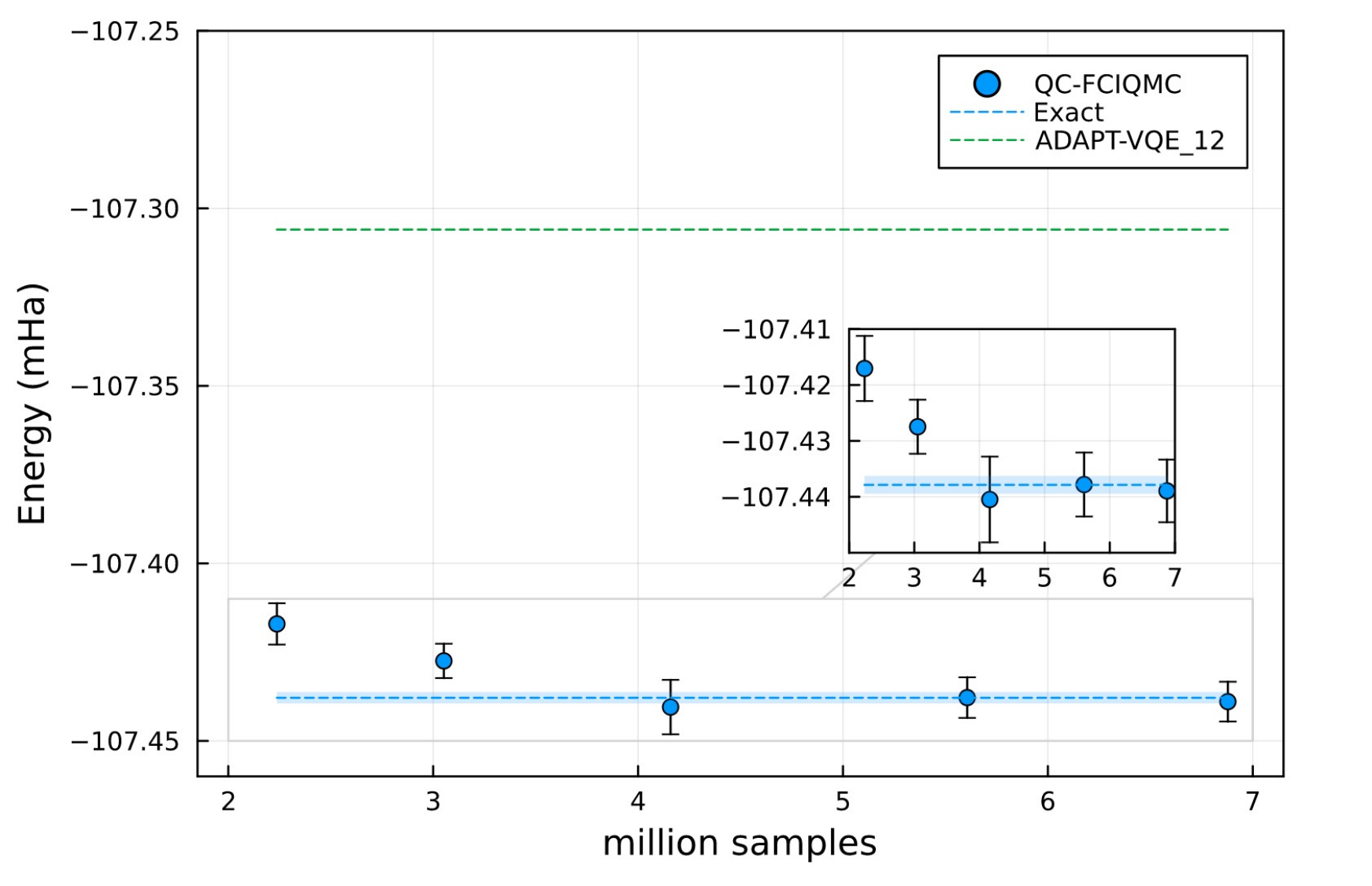}
    \caption{QC-FCIQMC on N$_2$ molecule at bond length 4.0$\AA$ when taking limited number of samples for the selection of $j$ configurations spawning from $i$. Here we use the VQE basis generated by 12 layers of ADAPT-VQE and the $x$-axis denotes the total number of samples of the whole QMC process. One can see the fewer samples we take, the larger the deviation from the exact energy. We can also observe that the variance does not vary much with the number of samples. Note that here the number of samples denotes the total number of samples needed for all configurations.}
    \label{sampling_noise}
\end{figure}

As we discussed in the last section, one can sample the configurations $j$s for which one calculates the corresponding matrix elements $H_{ij}$s with respect to a given $i$. One can choose a certain threshold for the number of $j$s given $i$ where one stops sampling more configurations. In Fig.~\ref{sampling_noise}, we plot the effects of taking a limited number of samples for the selection of $j$ configurations spawning from $i$ in N$_2$ simulation. We again use the VQE basis generated by 12 layers of ADAPT-VQE. We stop the sampling procedure after reaching a certain number of $j$ configurations for each $i$. We adjust this threshold to obtain the five data points in the figure and one can see the fewer samples we take, the larger the deviation from the exact energy. We can also observe that the variance does not vary much with the number of samples, which instead depends more on the number of layers of the ADAPT-VQE as shown in Fig.~\ref{N2_curve}(c).


\begin{figure}[h]
    \centering
    \includegraphics[height=8cm]{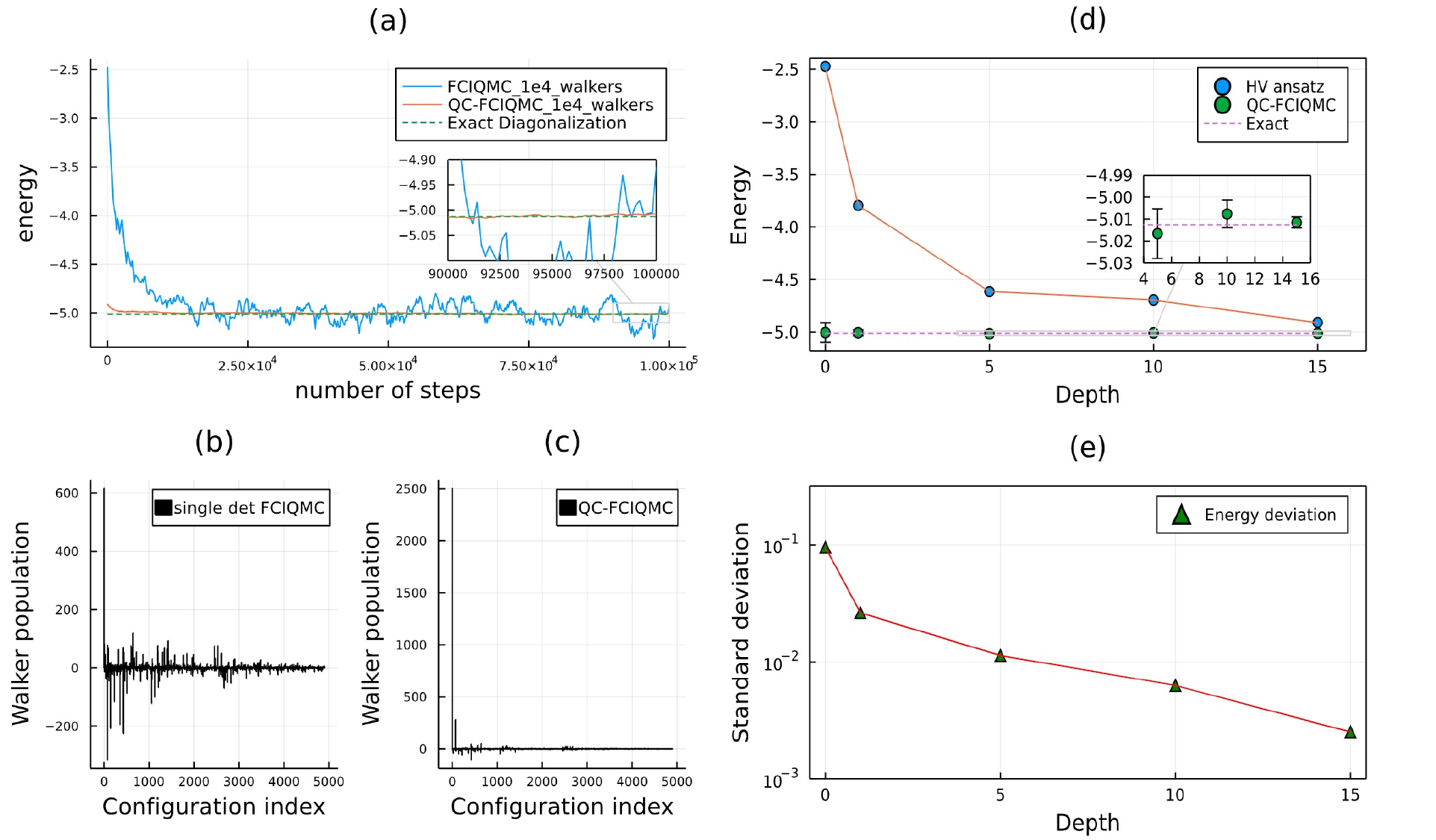}
    \caption{Comparison of the classical
     FCIQMC and QC-FCIQMC with basis generated by the Hamiltonian variational~(HV) ansatz on the $2\times 4$ Hubbard model~($U/t=4$). The HV ansatz with $1,5,10,15$ layers is implemented for comparison of variance suppression. The energy shift starts at $10^4$ walkers for all cases. (a) Comparison of energy fluctuation for FCIQMC and QC-FCIQMC. (b) Walker population for single determinant FCIQMC. (c) Walker population for QC-FCIQMC. (d) Energy estimation for different setups of layer number. (e) Effect of variance suppression.}
    \label{Hubbard}
\end{figure}

We also test our QC-FCIQMC algorithm on the Hubbard model. Here we choose the Hubbard model of size $2\time 4$, which requires 16 qubits and pick $U/t=4$ to enable stronger static correlation. We also choose the chemical potential as half-filling. In Fig.~\ref{Hubbard}, we compare the performance of QC-FCIQMC with the basis generated from the Hamiltonian variational ansatz~(HV ansatz)~\cite{wecker2015progress,cade2020strategies,cai2020resource} and the classical FCIQMC with single determinant basis. With the basis generated by a 15-layer HV ansatz, we observe the fluctuations of the energy much smaller in (a) and the walker population much more concentrated on a few configurations in (c) than the single-determinant basis in (b). We also plot the suppression of energy variance for HV ansatz with 1, 5, 10, and 15 layers in Fig.~\ref{Hubbard} (d) (e). We remark that the standard deviation of energy evaluation is reduced by almost one order of magnitude with merely a 1-layer HV ansatz.

\end{document}